\documentclass[twocolumn]{aastex631}
\usepackage{hyperref}
\usepackage{verbatim}
\usepackage{xcolor}
\usepackage{multirow}

\begin{document}

\shortauthors{Brodzeller et al.}
\shorttitle{DESI QSO Spectral Templates}

\title{Performance of the Quasar Spectral Templates for the Dark Energy Spectroscopic Instrument}

\author[0000-0002-8934-0954]{Allyson~Brodzeller}
\affiliation{Department of Physics and Astronomy, The University of Utah, 115 South 1400 East, Salt Lake City, UT 84112, USA}
\author{Kyle~Dawson}
\affiliation{Department of Physics and Astronomy, The University of Utah, 115 South 1400 East, Salt Lake City, UT 84112, USA}
\author[0000-0003-4162-6619]{Stephen~Bailey}
\affiliation{Lawrence Berkeley National Laboratory, 1 Cyclotron Road, Berkeley, CA 94720, USA}
\author{Jiaxi~Yu}
\affiliation{Ecole Polytechnique F\'{e}d\'{e}rale de Lausanne, CH-1015 Lausanne, Switzerland}
\author{A.~J.~Ross}
\affiliation{Center for Cosmology and AstroParticle Physics, The Ohio State University, 191 West Woodruff Avenue, Columbus, OH 43210, USA}
\affiliation{Department of Astronomy, The Ohio State University, 4055 McPherson Laboratory, 140 W 18th Avenue, Columbus, OH 43210, USA}
\affiliation{The Ohio State University, Columbus, 43210 OH, USA}
\author[0000-0002-9964-1005]{A.~Bault}
\affiliation{Department of Physics and Astronomy, University of California, Irvine, 92697, USA}
\author{S.~Filbert}
\affiliation{The Ohio State University, Columbus, 43210 OH, USA}


\author{J.~Aguilar}
\affiliation{Lawrence Berkeley National Laboratory, 1 Cyclotron Road, Berkeley, CA 94720, USA}
\author[0000-0001-6098-7247]{S.~Ahlen}
\affiliation{Physics Dept., Boston University, 590 Commonwealth Avenue, Boston, MA 02215, USA}
\author[0000-0002-5896-6313]{David~M.~Alexander}
\affiliation{Centre for Extragalactic Astronomy, Department of Physics, Durham University, South Road, Durham, DH1 3LE, UK}
\affiliation{Institute for Computational Cosmology, Department of Physics, Durham University, South Road, Durham DH1 3LE, UK}
\author{E.~Armengaud}
\affiliation{IRFU, CEA, Universit\'{e} Paris-Saclay, F-91191 Gif-sur-Yvette, France}
\author[0000-0003-3582-6649]{A.~Berti}
\affiliation{Department of Physics and Astronomy, The University of Utah, 115 South 1400 East, Salt Lake City, UT 84112, USA}
\author{D.~Brooks}
\affiliation{Department of Physics \& Astronomy, University College London, Gower Street, London, WC1E 6BT, UK}
\author[0000-0001-8996-4874]{E.~Chaussidon}
\affiliation{IRFU, CEA, Universit\'{e} Paris-Saclay, F-91191 Gif-sur-Yvette, France}
\author{A.~de la Macorra}
\affiliation{Instituto de F\'{\i}sica, Universidad Nacional Aut\'{o}noma de M\'{e}xico,  Cd. de M\'{e}xico  C.P. 04510,  M\'{e}xico}
\author{P.~Doel}
\affiliation{Department of Physics \& Astronomy, University College London, Gower Street, London, WC1E 6BT, UK}
\author[0000-0003-2371-3356]{K.~Fanning}
\affiliation{The Ohio State University, Columbus, 43210 OH, USA}
\author[0000-0003-1251-532X]{V.~A.~Fawcett}
\affiliation{School of Mathematics, Statistics and Physics, Newcastle University, Newcastle, UK}
\author[0000-0002-3033-7312]{A.~Font-Ribera}
\affiliation{Institut de F\'{i}sica d’Altes Energies (IFAE), The Barcelona Institute of Science and Technology, Campus UAB, 08193 Bellaterra Barcelona, Spain}
\author[0000-0003-3142-233X]{S.~Gontcho A Gontcho}
\affiliation{Lawrence Berkeley National Laboratory, 1 Cyclotron Road, Berkeley, CA 94720, USA}
\author{J.~Guy}
\affiliation{Lawrence Berkeley National Laboratory, 1 Cyclotron Road, Berkeley, CA 94720, USA}
\author{K.~Honscheid}
\affiliation{Center for Cosmology and AstroParticle Physics, The Ohio State University, 191 West Woodruff Avenue, Columbus, OH 43210, USA}
\affiliation{Department of Physics, The Ohio State University, 191 West Woodruff Avenue, Columbus, OH 43210, USA}
\affiliation{The Ohio State University, Columbus, 43210 OH, USA}
\author{S.~Juneau}
\affiliation{NSF's NOIRLab, 950 N. Cherry Ave., Tucson, AZ 85719, USA}
\author{R.~Kehoe}
\affiliation{Department of Physics, Southern Methodist University, 3215 Daniel Avenue, Dallas, TX 75275, USA}
\author[0000-0003-3510-7134]{T.~Kisner}
\affiliation{Lawrence Berkeley National Laboratory, 1 Cyclotron Road, Berkeley, CA 94720, USA}
\author[0000-0001-6356-7424]{Anthony~Kremin}
\affiliation{Lawrence Berkeley National Laboratory, 1 Cyclotron Road, Berkeley, CA 94720, USA}
\author[0000-0001-8857-7020]{Ting-Wen Lan}
\affiliation{Graduate Institute of Astrophysics and Department of Physics, National Taiwan University, No. 1, Sec. 4, Roosevelt Rd., Taipei 10617, Taiwan}
\author[0000-0003-1838-8528]{M.~Landriau}
\affiliation{Lawrence Berkeley National Laboratory, 1 Cyclotron Road, Berkeley, CA 94720, USA}
\author[0000-0003-1887-1018]{Michael E.~Levi}
\affiliation{Lawrence Berkeley National Laboratory, 1 Cyclotron Road, Berkeley, CA 94720, USA}
\author{C.~Magneville}
\affiliation{IRFU, CEA, Universit\'{e} Paris-Saclay, F-91191 Gif-sur-Yvette, France}
\author[0000-0002-4279-4182]{Paul~Martini}
\affiliation{Center for Cosmology and AstroParticle Physics, The Ohio State University, 191 West Woodruff Avenue, Columbus, OH 43210, USA}
\affiliation{Department of Astronomy, The Ohio State University, 4055 McPherson Laboratory, 140 W 18th Avenue, Columbus, OH 43210, USA}
\affiliation{The Ohio State University, Columbus, 43210 OH, USA}
\author[0000-0002-1125-7384]{Aaron M. Meisner}
\affiliation{NSF's NOIRLab, 950 N. Cherry Ave., Tucson, AZ 85719, USA}
\author{R.~Miquel}
\affiliation{Instituci\'{o} Catalana de Recerca i Estudis Avan\c{c}ats, Passeig de Llu\'{\i}s Companys, 23, 08010 Barcelona, Spain}
\affiliation{Institut de F\'{i}sica d’Altes Energies (IFAE), The Barcelona Institute of Science and Technology, Campus UAB, 08193 Bellaterra Barcelona, Spain}
\author[0000-0002-2733-4559]{J.~Moustakas}
\affiliation{Department of Physics and Astronomy, Siena College, 515 Loudon Road, Loudonville, NY 12211, USA}
\author[0000-0003-3188-784X]{N.~Palanque-Delabrouille}
\affiliation{IRFU, CEA, Universit\'{e} Paris-Saclay, F-91191 Gif-sur-Yvette, France}
\affiliation{Lawrence Berkeley National Laboratory, 1 Cyclotron Road, Berkeley, CA 94720, USA}
\author[0000-0002-0644-5727]{W.J.~Percival}
\affiliation{Department of Physics and Astronomy, University of Waterloo, 200 University Ave W, Waterloo, ON N2L 3G1, Canada}
\affiliation{Perimeter Institute for Theoretical Physics, 31 Caroline St. North, Waterloo, ON N2L 2Y5, Canada}
\affiliation{Waterloo Centre for Astrophysics, University of Waterloo, 200 University Ave W, Waterloo, ON N2L 3G1, Canada}
\author[0000-0001-7145-8674]{F.~Prada}
\affiliation{Instituto de Astrof\'{i}sica de Andaluc\'{i}a (CSIC), Glorieta de la Astronom\'{i}a, s/n, E-18008 Granada, Spain}
\author[0000-0002-3500-6635]{C.~Ravoux}
\affiliation{Aix Marseille Univ, CNRS/IN2P3, CPPM, Marseille, France}
\author{Graziano~Rossi}
\affiliation{Department of Physics and Astronomy, Sejong University, Seoul, 143-747, Korea}
\author[0000-0002-0408-5633]{C.~Saulder}
\affiliation{Korea Astronomy and Space Science Institute, 776, Daedeokdae-ro, Yuseong-gu, Daejeon 34055, Republic of Korea}
\author{M.~Siudek}
\affiliation{Institut de F\'{i}sica d’Altes Energies (IFAE), The Barcelona Institute of Science and Technology, Campus UAB, 08193 Bellaterra Barcelona, Spain}
\affiliation{Institute of Space Sciences, ICE-CSIC, Campus UAB, Carrer de Can Magrans s/n, 08913 Bellaterra, Barcelona, Spain}
\author[0000-0003-1704-0781]{Gregory~Tarl\'{e}}
\affiliation{University of Michigan, Ann Arbor, MI 48109, USA}
\author{B.~A.~Weaver}
\affiliation{NSF's NOIRLab, 950 N. Cherry Ave., Tucson, AZ 85719, USA}
\author[0000-0002-7520-5911]{S.~Youles}
\affiliation{Institute of Cosmology \& Gravitation, University of Portsmouth, Dennis Sciama Building, Portsmouth, PO1 3FX, UK}
\author[0000-0003-1887-6732]{Zheng~Zheng}
\affiliation{Department of Physics and Astronomy, The University of Utah, 115 South 1400 East, Salt Lake City, UT 84112, USA}
\author[0000-0001-5381-4372]{Rongpu Zhou}
\affiliation{Lawrence Berkeley National Laboratory, 1 Cyclotron Road, Berkeley, CA 94720, USA}
\author[0000-0002-4135-0977]{Zhimin~Zhou}
\affiliation{National Astronomical Observatories, Chinese Academy of Sciences, A20 Datun Rd., Chaoyang District, Beijing, 100012, P.R. China}

\correspondingauthor{Allyson Brodzeller}
\email{allyson.brodzeller@utah.edu}

\begin{abstract}
Millions of quasar spectra will be collected by the Dark Energy Spectroscopic Instrument (DESI), leading to
a four-fold increase in the number of known quasars. High accuracy quasar classification is essential to 
tighten constraints on cosmological parameters measured at the highest redshifts DESI observes ($z>2.0$). We 
present the spectral templates for identification and redshift estimation of quasars in the DESI Year 1 data 
release. The quasar templates are comprised of two quasar eigenspectra sets, trained on spectra from the Sloan
Digital Sky Survey. The sets are specialized to reconstruct quasar spectral variation observed over separate 
yet overlapping redshift ranges and, together, are capable of identifying DESI quasars from $0.05 < z <7.0$. 
The new quasar templates show significant improvement over the previous DESI quasar templates regarding 
catastrophic failure rates, redshift precision and accuracy, quasar completeness, and the contamination
fraction in the final quasar sample.
\end{abstract}

\section{Introduction}
\label{sect1:intro}

The Dark Energy Spectroscopic Instrument \citep[DESI;][]{DESIcollaboration16a,DESIcollaboration16b} will collect 
tens of millions of spectra from four extragalactic target classes: the bright galaxy sample
\citep[BGS;][]{hahn23}, luminous red galaxies \citep[LRG;][]{zhou23}, emission line galaxies 
\citep[ELG;][]{raichoor23}, and quasars \citep[QSO;][]{chaussidon23}. Roughly three million QSO spectra are 
expected over the five-year survey duration, increasing the number of spectroscopically confirmed QSOs by a 
factor of four \citep{lyke20}. 

Owing to their high intrinsic luminosity, QSOs are vital to understanding the large-scale structure of the 
universe by providing unparalleled access to the highest redshifts observed by DESI. QSOs can be used as direct 
tracers of the matter distribution and their spectra carry signatures of matter in the intergalactic medium that 
is otherwise undetectable. Measurements from QSOs complemented the low redshift galaxy tracers in the Baryon 
Oscillation Spectroscopic Survey and its extension \citep[BOSS, eBOSS;][]{dawson13,dawson16} to strengthen 
constraints on key cosmological parameters relating to expansion history, growth of structure, and inflation 
\citep{bourboux20,neveux20,hou21,mueller22}.

High accuracy, automatic classification of QSOs in DESI's larger spectroscopic sample is essential to make
successful measurements that use QSOs as either direct tracers or backlights to the neutral hydrogen of the 
Ly$\alpha$ forest. QSO misclassification dilutes the observed clustering amplitude while spectroscopic redshift
resolution can significantly impact the value of the $f\sigma_{8}$ parameter extracted from redshift space
distortions of the auto-correlation function \citep[e.g.][]{zarrouk18}. Additionally, random QSO redshift errors
dampen the radial peak of the baryon acoustic oscillation feature (BAO) and introduce unphysical correlations at
small transverse separations in the Ly$\alpha$-QSO cross-correlation measurement \citep{youles22}. 
Classification accuracy is complicated by the vast diversity of QSO spectral features. QSO emission lines 
exhibit varying degrees of asymmetry, broadening, and shifting dependent on an individual system's structure. 
Meanwhile, the continuum shape may be altered by both intrinsic and external absorption features. 

DESI spectra are automatically classified after collection into three broad categories: galaxy, QSO, or star, 
and simultaneously the redshifts are estimated. The most probable spectral class and redshift are determined by 
the 
Redrock\footnote{ \href{https://github.com/desihub/redrock/}{https://github.com/desihub/redrock/}}\textsuperscript{,}\footnote{The galaxy, previous QSO, and stellar 
spectral templates used in Redrock are available at
\href{https://github.com/desihub/redrock-templates/releases/tag/0.7.2}{https://github.com/desihub/redrock-templates/releases/tag/0.7.2}} 
\citep{bailey23} spectral template fitting algorithm. The QSO templates used for the DESI Survey Validation 
program \citep[SV;][]{DESIcollaboration23a} and the Early Data Release \citep[EDR;][]{DESIcollaboration23b} were 
adopted from (e)BOSS. The templates consist of four empirical eigenspectra derived via Principal Component 
Analysis (PCA) and described in Section~4.2 of \citet{bolton12}. These eigenspectra were determined from 568 QSO 
spectra from the Sloan Digital Sky Survey \citep[SDSS;][]{york00}. The small number statistics limit the range 
of spectral variation that can adequately be described. For example, the luminosity-dependent blue-shifting of 
high ionization lines \citep[e.g.][]{shen16} is not well incorporated into these templates. This feature 
contributes to redshift-dependent bias in QSO redshift estimates \citep[e.g.][]{chime23}. Further, QSO spectra 
exhibiting more peculiar spectral features, such as Broad Absorption Lines 
\citep[BAL; e.g.][]{hall02, gibson09, hall12, yi19} or reddened continua 
\citep[e.g.][]{glikman13,calistro21,fawcett22}, are often poorly modeled by the templates, leading to
misclassification in some cases.

Visual Inspection (VI) of early DESI data revealed high QSO sample purity and reliable redshift estimates with 
Redrock as the sole spectral classifier; however, approximately 11-15\% of true QSOs were misclassified as 
galaxies or stars  \citep{alexander23}. Two highly successful ``afterburner'' algorithms were implemented to 
retrieve the missed QSOs: QuasarNET \citep[QN;][]{busca18,farr20} and a broad Mg~\textsc{ii} emission line 
detection algorithm. QN is a convolutional neural network trained to identify QSO emission lines in spectra. 
Spectra missed by Redrock but identified as QSO by QN with a probability over a specified threshold are 
reclassified as QSO. The Redrock QSO templates are then refit to these spectra with a prior informed by the 
coarse QN redshift estimate to estimate final redshifts. The Mg~\textsc{ii} afterburner checks for significant 
Mg~\textsc{ii} emission, indicative of QSO activity, in spectra classified as `galaxy.' A QSO classification is 
assumed if detected, but the redshift estimate is unchanged. Together with Redrock, the afterburners brought the 
completeness up to approximately 94\%. A full description of afterburners, their performance, and the criteria 
for a spectrum to be labeled as QSO is presented by \citet{alexander23} and \citet{chaussidon23}.

This work presents new QSO templates developed for DESI and their performance. We expand upon the method 
presented by \citet[][hereafter B22]{brodzeller22} to create new QSO templates with a specific focus on spectral
diversity, redshift evolution, and redshift-dependent bias. Our method uses a clustering technique
(Section~\ref{subsect:clustering}) to compress a large sample of QSO spectra into subsets that, together, span
the range of the spectral variation. We use the compressed sample to remove contaminants such as misclassified
spectra and to standardize the rest-frame wavelength solution of the individual spectra. We then use the 
filtered and re-redshifted spectroscopic sample to compute eigenspectra that act as the QSO spectral templates 
for DESI.  

The most notable difference between the previous DESI QSO templates and the new DESI QSO templates is the use of 
two QSO template sets trained to different redshift ranges rather than one set covering the full redshift range.
We developed a low redshift and a high redshift QSO template set to allow for a more complete description of
spectral variation and redshift evolution over $0.05<z<7$, particularly in the extremes of this range. The low 
redshift templates classify QSO spectra over $0.05<z<1.6$ and are equipped to describe the host galaxy
contributions that are often more significant with lower redshift QSOs. The high redshift templates classify QSO
spectra over $1.4<z<7$ and are trained to reconstruct the variation observed in the highest luminosity systems.
The two template sets overlap in redshift coverage, as described in Section~\ref{subsect:eigenspectra}. These
templates are integrated into the Redrock software repository\footnote{\href{https://github.com/desihub/redrock-templates/releases/tag/0.8}{https://github.com/desihub/redrock-templates/releases/tag/0.8}} 
and will be used to produce the QSO classifications for the DESI Year 1 Data Release.

In Section~\ref{sect2:data}, we discuss the spectroscopic samples used to train the new QSO templates and to 
test their performance. We outline the template development method in Section~\ref{sect3:tempdev}, highlighting 
changes to the method presented by B22. In Section~\ref{sect4:perftest}, we present the performance of the new 
QSO templates on DESI spectra. We evaluate the results for Redrock alone versus Redrock together with the QSO 
afterburners. Finally, we measure the redshift accuracy of the new QSO templates through cross-correlations of 
QSOs with other tracers in DESI. We conclude in Section~\ref{sect5:summary} by summarizing this work and 
describing the potential for future studies. Throughout our work we assume flat $\Lambda$CDM cosmology described by \cite{planck16}.

\section{Data}
\label{sect2:data}
We use the wealth of QSO spectra from the Sixteenth Data Release \citep[DR16;][]{ahumada20} of the fourth
generation of the SDSS \citep[][]{blanton17} to develop QSO spectral templates. The classification 
performance of the new QSO templates in Redrock alone and in the Redrock with QSO afterburners pipeline is
tested on subsets of DESI spectra collected between December 2020 and July 2021. We compare the results to
those from the previous DESI QSO templates on the same samples. In this section, we review the SDSS training
sample and DESI validation samples.

\subsection{Training Sample}
\label{subsect:trainingsample}

We select spectra from SDSS DR16 using attributes from the DR16 QSO catalog \citep{lyke20}. The spectra were
collected with the Sloan telescope \citep{gunn06} using the (e)BOSS spectrographs \citep{smee13}. QSO targets
in this analysis were selected as described in \citet{ross12} and \citet{myers15}. 

We use QSO spectra with a \verb|Z_PCA| redshift of $0.05 \leq z \leq 5.0$ and \verb|ZWARN_PCA|~$=0$ to develop
the new QSO templates. The method for estimating \verb|Z_PCA| is explained in Section~4.4 of \citet{lyke20}. 
Pixels in SDSS spectra are binned on a logarithmic wavelength array with constant spacing. The average S/N per
pixel of each spectrum is determined over all pixels with a rest-frame (RF) wavelength of 
$\lambda_{RF} > 1216$~\AA, after excluding pixels flagged as having poor sky subtraction. We also exclude
pixels at wavelengths shorter than 1216~\AA\ from the S/N measurement, as Ly$\alpha$ absorption features 
dominate the spectrum at these wavelengths. Spectra with an average S/N per pixel~$>5$ are included in the
training sample. The final selection provides 207,956 unique spectra.

We reduce the impact of non-intrinsic foreground features in the spectra by applying various corrections. The 
corrections applied are identical to that in B22 except with changes to filtering outlier pixels. We first
correct for galactic extinction using the dust map of \citet{schlegel98} and extinction law of 
\citet{fitzpatrick99}. We then mask pixels flagged for poor sky subtraction and those that deviate negatively
by more than $2\sigma$ from the ratio of measured flux to flux smoothed by a 30-pixel median filter 
(corresponding to approximately 2000~km~s$^{-1}$). The latter mask aims to remove contamination from foreground
IGM absorption but may unintentionally mask some intrinsic absorption features, such as narrow BALs. Lastly, each 
spectrum is shifted to its rest-frame solution.

\subsection{Validation Samples}
\label{subsect:validationsample}

DESI is installed on the Mayall Telescope at Kitt Peak National Observatory. The focal plane holds 5,000 optical
fibers, steered by robotic positioners to the locations of survey targets \citep{myers23,silber23}. The focal
plane is split into ten wedges, called petals, each holding 500 fibers that direct light to a multi-object fiber 
spectrograph. A single pointing of the ten petals is collectively referred to as a tile. A complete description
of DESI instrumentation is provided in \citet{DESIcollaboration22}.

Spectra collected by DESI are processed in a fully automatic pipeline \citep{guy23} and then classified by 
Redrock and the QSO afterburners. The spectra are binned on a wavelength array with constant linear spacing 
between pixels. 

We use five samples of DESI spectra for validating the new QSO templates, detailed in the following paragraphs:
The DESI Truth Table, QSO Main, Survey Validation (SV) Repeat Exposures, the One-Percent Survey, and the 
Guadalupe Internal Data Assembly (hereafter Guadalupe). All spectra were classified with Redrock version~0.15.4
and the QSO afterburners using the previous QSO templates. The SV and One-Percent spectra and corresponding 
catalogs with these classifications will be publicly available in the EDR. The Guadalupe data used in this work
will be released as part of DESI Year 1 and will use classifications from the QSO templates presented in this 
paper.

\textbf{DESI Truth Table:} The first validation sample is selected from the SV program and consists of spectra
from eight tiles: 80605, 80606, 80607, 80608, 80609, 80610, 80613, and 80742. These tiles were exposed multiple
times to achieve a higher cumulative effective exposure depth than the nominal main survey. Coadding all the
spectra obtained for a target enabled reliable VI of early survey data to assess pipeline performance and
preliminary target selection methods \citep{allendoprieto20,raichoor20,ruizmacias20,yeche20,zhou20}. As a 
result, there exist VI labels for many of these spectra that include a spectral class, redshift, and confidence
in the visual assessment. Confidence is provided on a scale of 0 to 4, with 0 indicating no signal and 4
indicating two or more secure spectral features are identifiable. A VI confidence of equal to or greater than 
2.5, meaning at least one inspector indicated a probable classification with at least one clear spectral feature
or several weak features, is treated as truth \citep[see][for details]{alexander23,lan23}. 

We use the truth labels to determine the classification accuracy and sample completeness with the new QSO 
templates for the resulting QSO and galaxy samples (Section~\ref{subsect:classification_accuracy}). In total,
we test performance on 17,397 cumulative spectra with confident VI labels. The breakdown of spectral classes
assigned from VI in the DESI Truth Table is given Table~\ref{table_VI}.

\begin{table}[htb]
\centering
\caption{VI spectral class of objects in the DESI Truth Table (All Targets) and the QSO Main subset.}

\begin{tabular}{c | c c } 

\hline
Spectral & \multicolumn{2}{c}{Number of Spectra} \\
Class    & \emph{All Targets}       & \emph{QSO Main} \\
 \hline
QSO & 1643 & 1368 \\
Galaxy & 14991 & 915 \\
Star & 763 & 349 \\ 

\hline
\end{tabular}
\label{table_VI}
\end{table}

\textbf{QSO Main:} The QSO target selection for SV includes fainter objects than the main survey
\citep{chaussidon23}. We study performance on main survey QSO targets by restricting the DESI Truth Table to
target type `QSO' and $r < 23$. This subsample contains 2,632 cumulative spectra, of which a non-negligible 
fraction is non-QSO according to VI. The VI spectral class distribution within QSO Main is included in
Table~\ref{table_VI}. 

\textbf{SV Repeat Exposures:} The individual exposures of targets in the DESI Truth Table tiles and six 
additional deep tiles (80620, 80678, 80690, 80695, 80699, 80711) were randomly sampled to create several spectra
of each target equivalent to exposures at main survey depth. We use the spectra of QSO targets in this sample to
quantify the redshift precision by comparing the redshift estimates for different spectra of the same target
(Section~\ref{subsubsect:repeat_exposures}). 

\textbf{One-Percent Survey and Guadalupe:} The One-Percent Survey, occasionally referred to as SV3
\citep[][]{chaussidon23}, was collected primarily between April 5, 2021 and May 13, 2021 using exposure times
20\% longer than the fiducial. Guadalupe is the first two months of main survey data collected by DESI between
May 14th and July 9th of 2021. We use these last two samples for measuring the bias on QSO redshifts and for
additional testing of QSO redshift precision (Section~\ref{subsubsect:cross-correlation}). We also use Guadalupe
to check the spectra of targets that change classification with the change of QSO templates
(Section~\ref{subsubsect:classification_accuracy_MS}), evaluating the overall accuracy of the changes and
determining any new potential failure modes.

\section{QSO Template Development}
\label{sect3:tempdev}

\begin{figure}[htb]
  \centering
	  \includegraphics[width=0.25\textwidth, angle=0]{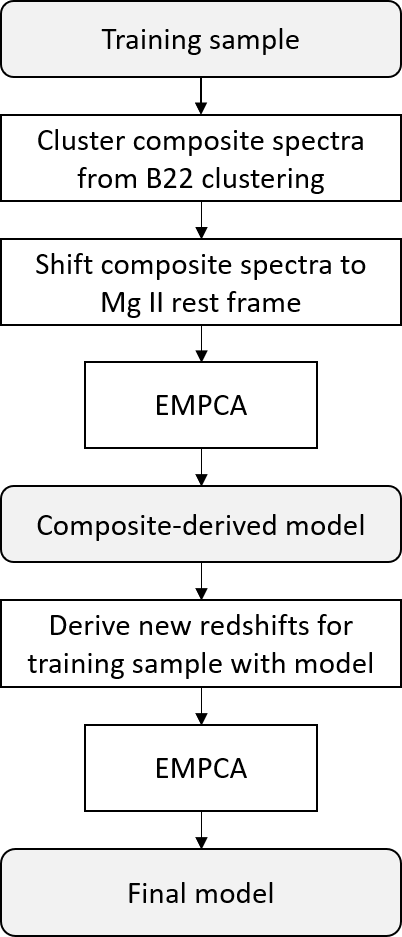}
	  \caption{Flowchart of template development process.}
	  \label{fig:fig1}
\end{figure}

This section provides a brief overview of the clustering technique and the compression of the resulting clusters
into a PCA training set of composite spectra. We primarily focus on the changes from B22, referring the reader
there for complete details. We then discuss the process for re-redshifting the training sample which is also 
used to prune noisy, misclassified, or problematic spectra from the training sample. Finally, we present the
final PCA eigenspectra sets that comprise the QSO templates for classification in DESI. The overall process for template development is outlined in Figure~\ref{fig:fig1}.

\subsection{Overview of the Clustering Procedure and Results}
\label{subsect:clustering}

We use the clustering technique of B22 to compress the training sample of QSO spectra into groups that exhibit
low internal diversity. The clustering method is based on 
SetCoverPy\footnote{\href{https://github.com/guangtunbenzhu/SetCoverPy}{https://github.com/guangtunbenzhu/SetCoverPy}},
an archetype technique developed by \citet{zhu16}. The only free parameter in the clustering method is the 
threshold reduced $\chi^2$ (B22 Equation~1), below which two spectra are considered similar and therefore belong 
to the same cluster. This parameter is tuned to maximize the similarity of spectra in the same cluster without 
creating excessive outliers (i.e., spectra not within the threshold $\chi^2_{red}$ to any other spectrum). We
refer to Section~3 of B22 for a full description of the clustering technique and outlier identification. We only
provide a brief overview, focusing on details specific to this work.

\begin{table*}[htb]
\centering
\caption{The number of QSOs in each redshift bin, the median S/N of spectra in the bin, and rest-frame 
wavelength range used for clustering. The median S/N is used to split the redshift bin into two evenly sized 
sub-bins in which clustering is performed independently. The final number of clusters and outliers are reported 
after the outlier retrieval stage. The lowest redshift bin is not split on S/N owing to small numbers and the 
results are for the full bin. A different clustering method is used for the highest redshift 
bin, described in the text.}

\centerline{
\tablewidth{0pt}
\hskip-2.5cm\begin{tabular}{c | c c c c | c c c | c c c}

\hline
         &            &            &                 &                 &         \multicolumn{3}{c|}{High S/N}          &       \multicolumn{3}{c}{Low S/N}            \\
\hline
\multirow{2}{*}{Redshift} & \multirow{2}{*}{QSOs} & Median & $\lambda_{min}$ & $\lambda_{max}$ & \multirow{2}{*}{Initial $\chi^{2}_{red}$} & \multirow{2}{*}{Clusters} & \multirow{2}{*}{Outliers} & \multirow{2}{*}{Initial $\chi^{2}_{red}$} & \multirow{2}{*}{Clusters} & \multirow{2}{*}{Outliers} \\ 
            &         & S/N  & [\AA] & [\AA] &  &   &   &  &   &   \\
 \hline
$0.0<z<0.35$ & 3,289  &  -   & 3611 & 7687 & 2.477 & 138 & 259 &   -   &  -  &  -  \\ 
$0.35<z<0.7$ & 15,847 & 7.89 & 2674 & 6104 & 1.320 & 451 & 305 & 1.053 & 464 & 228 \\
$0.7<z<0.85$ & 11,541 & 7.45 & 2124 & 5617 & 1.241 & 362 & 140 & 1.067 & 390 & 82  \\
$0.85<z<1.0$ & 12,125 & 7.47 & 1952 & 5189 & 1.100 & 411 & 102 & 0.989 & 392 & 55  \\ 
$1.0<z<1.15$ & 12,218 & 7.49 & 1805 & 4818 & 1.034 & 335 & 43  & 0.950 & 296 & 29  \\
$1.15<z<1.3$ & 13,048 & 7.42 & 1679 & 4519 & 1.016 & 337 & 58  & 0.950 & 330 & 51  \\

\hline
$1.0<z<1.2$ & 16,463 & 7.47 & 1805 & 4724 & 1.083 & 472 & 66  & 0.950 & 496 & 61  \\ 
$1.2<z<1.4$ & 18,117 & 7.43 & 1641 & 4332 & 1.080 & 445 & 83  & 0.952 & 489 & 57  \\ 
$1.4<z<1.6$ & 17,992 & 7.34 & 1505 & 3999 & 1.149 & 480 & 167 & 1.017 & 527 & 83  \\ 
$1.6<z<1.8$ & 16,918 & 7.36 & 1389 & 3714 & 1.266 & 432 & 300 & 1.083 & 473 & 109 \\ 
$1.8<z<2.0$ & 17,310 & 7.76 & 1289 & 3460 & 1.376 & 398 & 510 & 1.123 & 438 & 154 \\ 
$2.0<z<2.2$ & 17,145 & 7.71 & 1203 & 3249 & 1.609 & 369 & 569 & 1.255 & 399 & 242 \\ 
$2.2<z<2.4$ & 20,155 & 7.60 & 1128 & 3058 & 1.899 & 435 & 670 & 1.441 & 531 & 296 \\ 
$2.4<z<2.6$ & 14,796 & 7.60 & 1062 & 2889 & 2.077 & 330 & 602 & 1.516 & 423 & 344 \\
\hline
$2.6<z<5.0$ & 26,258 &  -   &   -  &  -   &   -   & -   & 616 & -     & -   & 464  \\
\hline
 
\end{tabular}
}

\label{table_clustering}
\end{table*}

We use the $\chi^{2}_{red}$ distance metric from B22 to measure the similarity between all pairs of spectra 
within a clustering bin (bins are described in the following paragraphs). The $\chi_{red}^2$ metric includes a 
two-parameter term ($c\lambda^{\Delta\alpha}$) that is applied to the higher S/N spectrum of each pair to 
account for the difference in broadband continuum between the spectra. The values of $c$ and $\Delta\alpha$ are 
determined for each pair so that the $\chi^{2}_{red}$ is minimized. QSO continua are well described by a single 
power-law between the Ly$\alpha$ and H$\beta$ emission lines \citep[e.g.][]{brotherton01,vandenberk01}, 
permitting this term to be reasonably applied to QSO pairs at $z>1$. The term's validity breaks at lower 
redshifts as wavelengths longer than that of the H$\beta$ line enter the spectral coverage. We therefore split 
the training sample into two overlapping subsamples for clustering:\footnote{note that this division of the 
training sample is not the same that is used to develop the final QSO templates in this work, see 
Section~\ref{subsect:eigenspectra}} $1.0 < z < 2.6$ and $z < 1.3$. For the $z<1.3$ subsample, $c=1$ and 
$\Delta\alpha = 0$ in all $\chi^{2}_{red}$ calculations. Spectra with redshifts above $z=2.6$ are excluded from 
initial clustering and incorporated into clusters at a subsequent stage, as described later in this section.

The training subsamples are binned on discrete intervals of $\Delta z \approx 0.15 - 0.35$ to account for 
redshift evolution of spectral features, described in Table~\ref{table_clustering}. Within each redshift bin, 
the spectra are cropped to the overlapping wavelength range defined by the shortest wavelength of the lowest 
redshift spectrum and the longest wavelength of the highest redshift spectrum in the bin. Additionally, pixels 
with a rest-frame wavelength shorter than 1216~\AA\ are masked to avoid spurious signal from the Ly$\alpha$ 
forest. The spectra are then normalized according to the median flux density measured over the common 
rest-frame wavelength range. The pixel diversity mask (B22 Section~3.2.1) is applied at this stage. This mask 
focuses spectral comparisons on the top 32\% of pixels exhibiting the most flux variation within a given 
redshift bin to maximize the information content in the resulting clusters.

Each redshift bin is further binned on its median S/N, excluding the lowest redshift bin owing to low numbers.
Our motivation for subbinning on S/N is two-fold. First, the noise levels vary greatly across the sample and 
splitting on S/N allows us to tune the threshold distance for clustering to achieve a meaningful measure of 
similarity between spectra given noise levels. Second, many spectral features are known to evolve with
luminosity \citep[e.g.][]{shields07,jensen16}. We can treat S/N as a loose proxy for luminosity since the QSOs
are at the same approximate redshift and SDSS exposure times are relatively consistent.

Clustering is performed independently in the high S/N and low S/N bins of each redshift bin. We enforce
$\chi^{2}_{red} \geq 0.95$ when determining the initial clustering threshold distances. The same outlier
retrieval method in B22 is applied to the resulting clusters to maximize diversity coverage. The initial
threshold $\chi^{2}_{red}$ and final clustering results are summarized in Table~\ref{table_clustering}.

The clustering technique produces 11,043 non-exclusive clusters of similar spectra and 6,388 unique outliers. 
Cluster size ranges from two spectra to several hundred spectra. The smaller clusters often contain more
peculiar QSOs such as BALQSOs, dust-reddened QSOs, systems with extremely shifted lines, or QSOs with unusually
high equivalent widths while larger clusters contain more typical QSO spectra displaying blue continua. Though
the degree of overlap between clusters varies greatly, larger clusters tend to share more spectra with other
larger clusters than smaller clusters do in general. 

To first order, the collection of the clusters represents the range of spectral diversity observed across the 
full training sample. Many spectra identified as outliers in clustering exhibit calibration or classification
errors, peculiar features such as BALs or atypical emission line strengths or ratios, or have incorrect
redshifts from \verb|Z_PCA|, as determined from VI. The natural segregation of these spectra facilitate
contaminant rejection before training the final spectral model (See 
Section~\ref{subsect:redshiftcorrection_comps}).

The $\chi^{2}_{red}$ used to cluster QSO spectra excludes pixels in the Ly$\alpha$ forest. Above $z=2.6$, the
Ly$\alpha$ forest is present in an increasingly significant portion of the spectrum. For this reason, QSOs at 
$2.6<z<5.0$ have been excluded from the first stage of clustering. We retrieve these high redshift spectra by 
adding them to the existing clusters of the $2.4<z<2.6$ bin in a similar fashion to the outliers.

We first divide the $z>2.6$ spectra on the median S/N of the $2.4<z<2.6$ redshift bin. The $\chi^{2}_{red}$ is
calculated between pairs of the $z>2.6$ spectra and cluster representative spectra in the S/N bin corresponding
to the high redshift spectrum's S/N. The distance metric is determined in the overlapping wavelength range of 
the pair, restricting the comparison to $1216$~\AA\ ~$< \lambda_{RF} < 2884$~\AA. A high redshift spectrum is
then added to all clusters for which the $\chi^{2}_{red}$ to the representative spectrum is less than 150\% of
the initial distance threshold of that bin. 

An error-weighted mean composite spectrum is then computed from all clusters as follows. The processed spectra
(Section~\ref{subsect:trainingsample}) belonging to clusters from the $z<1.3$ subsample are normalized over
$3620-4510$~\AA\ in the rest frame. The processed spectra belonging to clusters from the $1.0<z<5.0$ subsample, 
excluding $z>2.4$, are normalized over $1815-2880$~\AA. A different rest-frame range of $1275-1725$~\AA\ is
required to normalize the spectra in clusters at $z>2.4$ owing to the extended wavelength coverage of the
clusters. 

All individual spectra are adjusted to the continuum slope of the representative spectrum using the 
$c\lambda^{\Delta\alpha}$ correction when taking the error-weighted mean of a cluster in the $1.0<z<5.0$ 
subsample. The cluster composite spectra from the $z<1.3$ subsample are simple error-weighted means. Thirty 
pixels on each extreme of a composite spectrum are removed to reduce edge noise. Lastly, we mask missing flux 
values and outlier pixels in each composite spectrum. Outlier pixels are identified identically as in the 
training sample, except the threshold to be masked is $3\sigma$ positive or negative.

\subsection{Correcting Redshifts in the Training Sample}
\label{subsect:redshiftcorrection_comps}

We reported a redshift-dependent bias on \verb|Z_PCA| in B22. We showed the bias could be stabilized by
eigenspectra trained in the rest frame of the Mg~\textsc{ii} emission line (see B22 Figure~9). The use of
Mg~\textsc{ii} as a stable redshift estimator is similarly reported by \citet{hewett10}. We therefore shift 
all composite spectra so that the Mg~\textsc{ii} line is at the expected rest-frame wavelength using an 
improved version of the redshift correction method in B22 (see Section~3.3 of B22). This correction is enabled
by the high S/N of the composite spectra, which facilitates a direct fit to the line location.

We approximate the underlying flux that is not associated with Mg~\textsc{ii} emission by fitting a power-law
over $2600-2725$~\AA\ and $2875-2950$~\AA. After subtracting the power-law fit, a double Gaussian is fit to 
$2730-2870$~\AA. We impose a lower bound on the dispersion of 250~km~s$^{-1}$ for both components of the 
Gaussian and require the two peak positions to be identical. 
The continuum and line profile are fit twice with pixels that deviate by more than $3\sigma$ from the first fit
masked in the second. The iterative fitting reduces the influence of associated absorption features on 
estimating the line location. The wavelength solution of the spectrum is then shifted so the line center is at 
the rest-frame wavelength of 2798.75~\AA\ \citep[consistent with][]{vandenberk01}. This method for fitting the
continuum and line profile neglects Fe~\textsc{ii} emission \citep[e.g.][]{wills80,wills85,vestergaard01} which 
can impact measured line widths. Since we are only concerned with determining the line center, we do not 
include an iron template and rely on the continuum estimate to absorb iron contributions. Systematic offsets on 
our Mg~\textsc{ii} location estimate will contribute to redshift error on the final QSO templates for DESI. We 
explore the magnitude of systematic redshift offsets in Section~\ref{subsubsect:cross-correlation}.

Cluster composite spectra of the 10,905 clusters from $z>0.35$ redshift-S/N subbins 
(Table~\ref{table_clustering}) are corrected with this method. The procedure fails to fit the Mg~\textsc{ii}
emission line in 134 cases. VI of these spectra reveals clear contamination from stellar spectra or redshift
errors in most cases. We reject these composite spectra and their contributing spectra from further analyses.

The Mg~\textsc{ii} emission line falls off the blue end of (e)BOSS spectra at $z\leq0.35$. It therefore cannot
be used to correct for potential redshift offsets in the composite spectra of $z<0.35$ clusters. We use a 
different approach to place these composite spectra in the same frame as the others. We apply Weighted
Expectation Maximization Principal Component Analysis
\citep[EMPCA;\footnote{\href{https://github.com/sbailey/empca/}{https://github.com/sbailey/empca/}}][]{bailey12}
to the mean-subtracted re-redshifted cluster composite spectra of the $z<1.3$ subsample. EMPCA was designed to
determine eigenvectors of data sets with noise or gaps, using measurement uncertainty to weight input data. The
composite spectra are median normalized over their common wavelength coverage and weighted by their propagated
inverse variance. We then fit the mean spectrum plus the first five eigenspectra from EMPCA to the composite
spectra of the $z<0.35$ clusters. We fit only $3100-7000$~\AA\ by the current rest-frame solution to ensure
coverage by the eigenspectra. Each spectrum is fit at intervals of 69~km~s$^{-1}$, the resolution of (e)BOSS 
spectra, over $\pm 2070$~km~s$^{-1}$ and then shifted to the new redshift solution.

\begin{figure}
  \centering
	  \includegraphics[width=0.45\textwidth, angle=0]{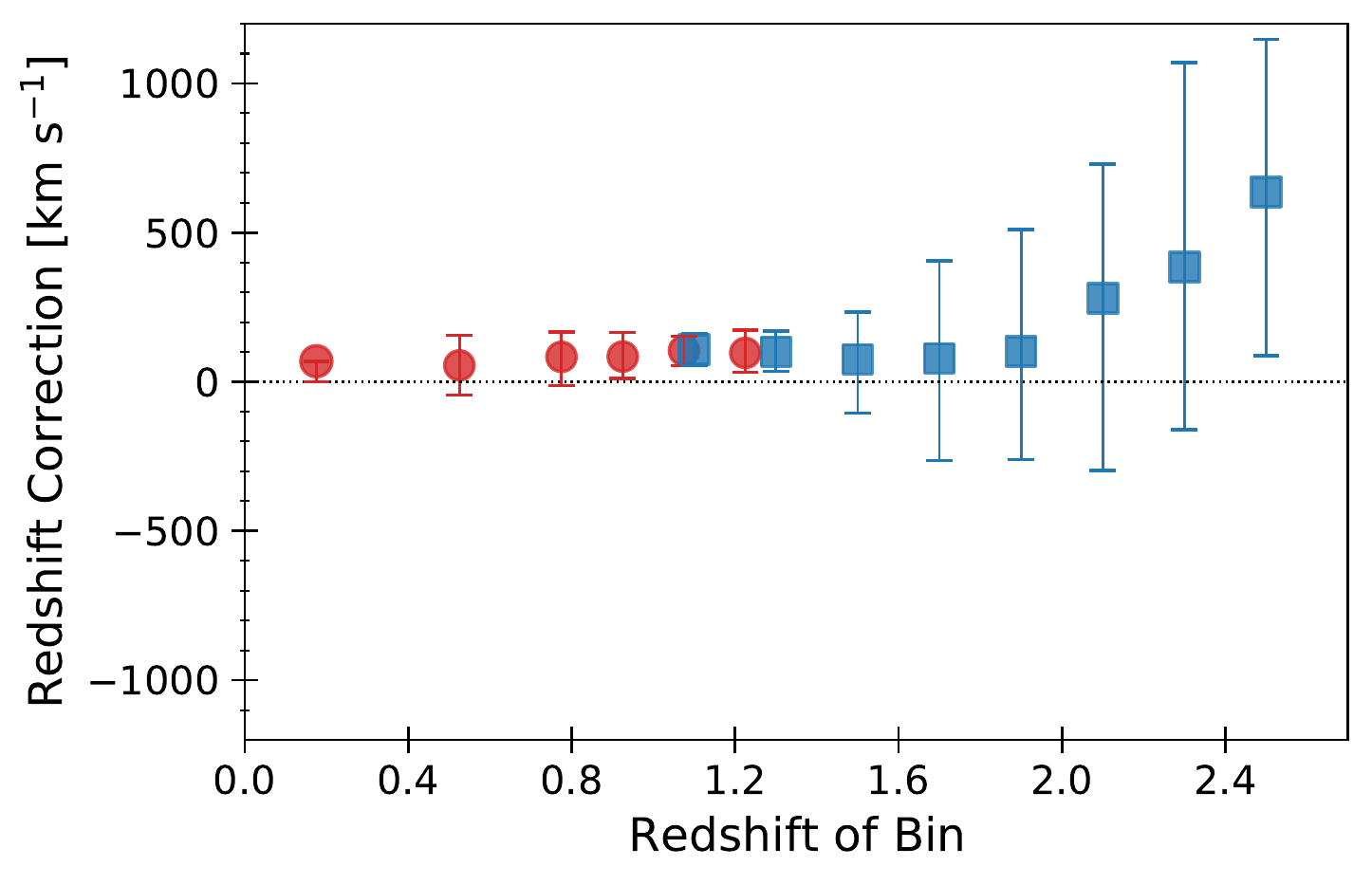}
	  \caption{Red circles (blue squares) indicate the median redshift correction applied to cluster composite
	  spectra of each low (high) redshift bin, including the $z<0.35$ bin. Error bars indicate the 16th and
	  84th percentiles of each distribution.}
	  \label{fig:fig2}
\end{figure}

As shown in Figure~\ref{fig:fig2}, larger redshift corrections are required with increasing redshift and the 
corrections tend more red (\verb|Z_PCA| is underestimated relative to the Mg~\textsc{ii} rest frame). The trend
is possibly driven by the blue-shifting of high ionization lines, such as C~\textsc{iv} and Ly$\alpha$, with 
respect to Mg~\textsc{ii}, a less biased line. The greater flux density of these lines relative to 
Mg~\textsc{ii} tends to dominate the fit during redshift estimation. Further, the Ly$\alpha$ forest is present 
in spectra at $z>1.92$, a clear inflection point in this figure. 

After redshift corrections, we subtract the error-weighted mean of cluster composite spectra of the $z<1.3$
subsample from all composite spectra within that subsample. We then derive eigenspectra from the residuals 
using EMPCA. We weight the flux values by the propagated inverse variance. This results in a model consisting 
of a mean vector and the first five eigenspectra from EMPCA that covers approximately $1800 - 9600$~\AA\ in the
rest frame. We repeat these steps with the cluster composite spectra from the $1.0<z<5.0$ subsample, creating a
second spectral model with different rest-frame wavelength coverage of approximately $600 - 5200$~\AA. The high
redshift model is padded with an exponential decay out to $\lambda_{RF} = 590$~\AA\ to provide redshift 
coverage up to $z=5.05$.

New to B22, these eigenspectra do not comprise our final QSO templates. There are two main issues with using 
the composite-derived templates for classification in DESI. First, all the spectra within a cluster have 
individual redshift errors. Although small, the varying redshift offsets within a cluster lead to blurring and 
peak dampening of the emission lines in the composite spectrum. These features may negatively affect modeling. 
Second, the clusters overlap in membership with one spectrum potentially belonging to tens of clusters. The 
cluster overlap is desirable to avoid harsh boundaries on smooth population gradients, but, if we use 
eigenspectra from the cluster composites as the QSO templates, we introduce an artificial up-weighting of more 
common spectral features.  

We instead use Redrock with the high and low redshift QSO models derived from the composite spectra to fit the 
calibrated spectra of the training sample and derive new redshifts. We mask pixels in the \verb|Z_PCA| rest 
frame at $\lambda_{RF} < 1216$~\AA\ while fitting. Additionally, if the spectrum has a non-zero \verb|BAL_PROB|
attribute indicating BALs are present \citep{guo19}, we mask pixels in the following \verb|Z_PCA| rest-frame 
ranges: $1479-1549$~\AA, $1326-1397$~\AA, and $1216-1240$~\AA. These ranges correspond to the typical 
wavelengths of C~\textsc{iv}, Si~\textsc{iv} and N~\textsc{v} BALs. Lastly, we also remove pixels at 
$\lambda_{obs} > 9700$~\AA\ owing to lack of coverage from the low redshift templates as they approach $z 
\approx 0.05$. This re-redshifting step aims to reduce the redshift bias seen in Figure~\ref{fig:fig2} while 
addressing the issues outlined in the previous paragraph.

In the overlap redshift range of the models, we observe strong agreement in redshift estimates from the two 
eigenspectra sets. For these spectra, we use the redshift from the model that provided the lowest 
$\chi^2_{red}$ fit. We reject spectra from the final training sample that have a non-zero \verb|ZWARN| 
attribute or $|\Delta z| > 0.05$ between the new redshift and \verb|Z_PCA|. This cut effectively removes all 
the outlier spectra that exhibit calibration errors, incorrect classifications, or little signal and leaves 
most true QSO systems in the final sample except those displaying extreme peculiarities. We refine the redshifts 
of spectra with a new redshift of $z>1.92$ by updating the Ly$\alpha$ forest mask and refitting over 
$\pm$~5000~km~s$^{-1}$ from the initial redshift estimate. On average, the initial redshift estimate with these 
templates is redder than \verb|Z_PCA| and the refined redshifts are almost always redder than the initial 
redshift. Our final model is trained using these new redshifts.

\subsection{The Eigenspectra}
\label{subsect:eigenspectra}

We suggested in B22 that using multiple QSO template sets, each trained to reconstruct the variation observed
over limited redshift intervals, should improve spectral modeling over the full QSO population
\citep[supported by][]{yip04}. Toward this goal, we create two QSO eigenspectra sets that together serve as 
the QSO templates for spectral classification in DESI. The new DESI QSO templates consist of a low redshift 
eigenspectra set and a high redshift eigenspectra set. The sets are trained on subsamples of the individual 
spectra of the re-redshifted, pruned training sample introduced in the previous subsection. The templates
are derived with EMPCA, weighting flux values by their inverse variance.

\begin{figure*}
  \centering
	  \includegraphics[width=0.48\textwidth, angle=0]{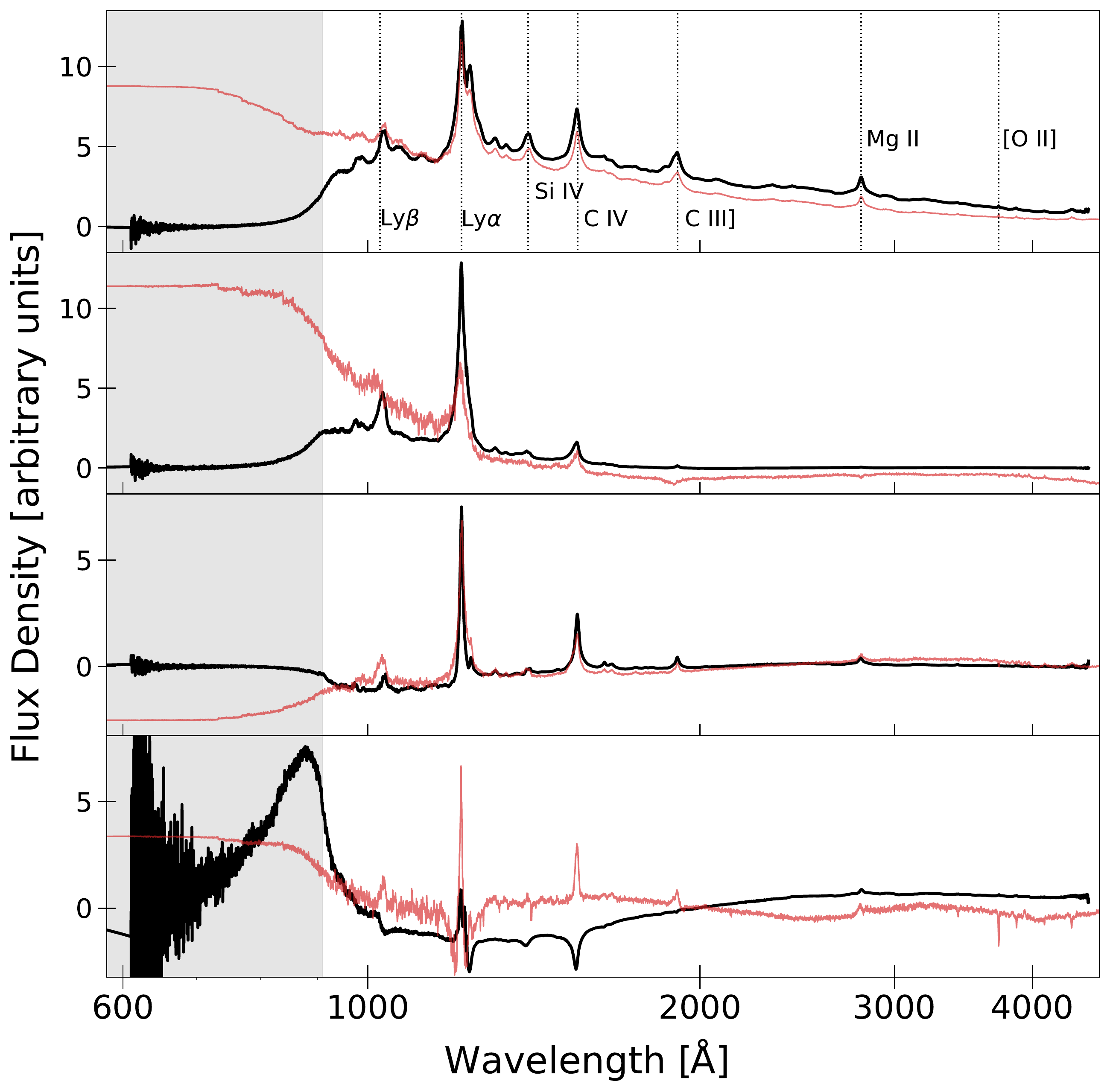}
	  \includegraphics[width=0.48\textwidth, angle=0]{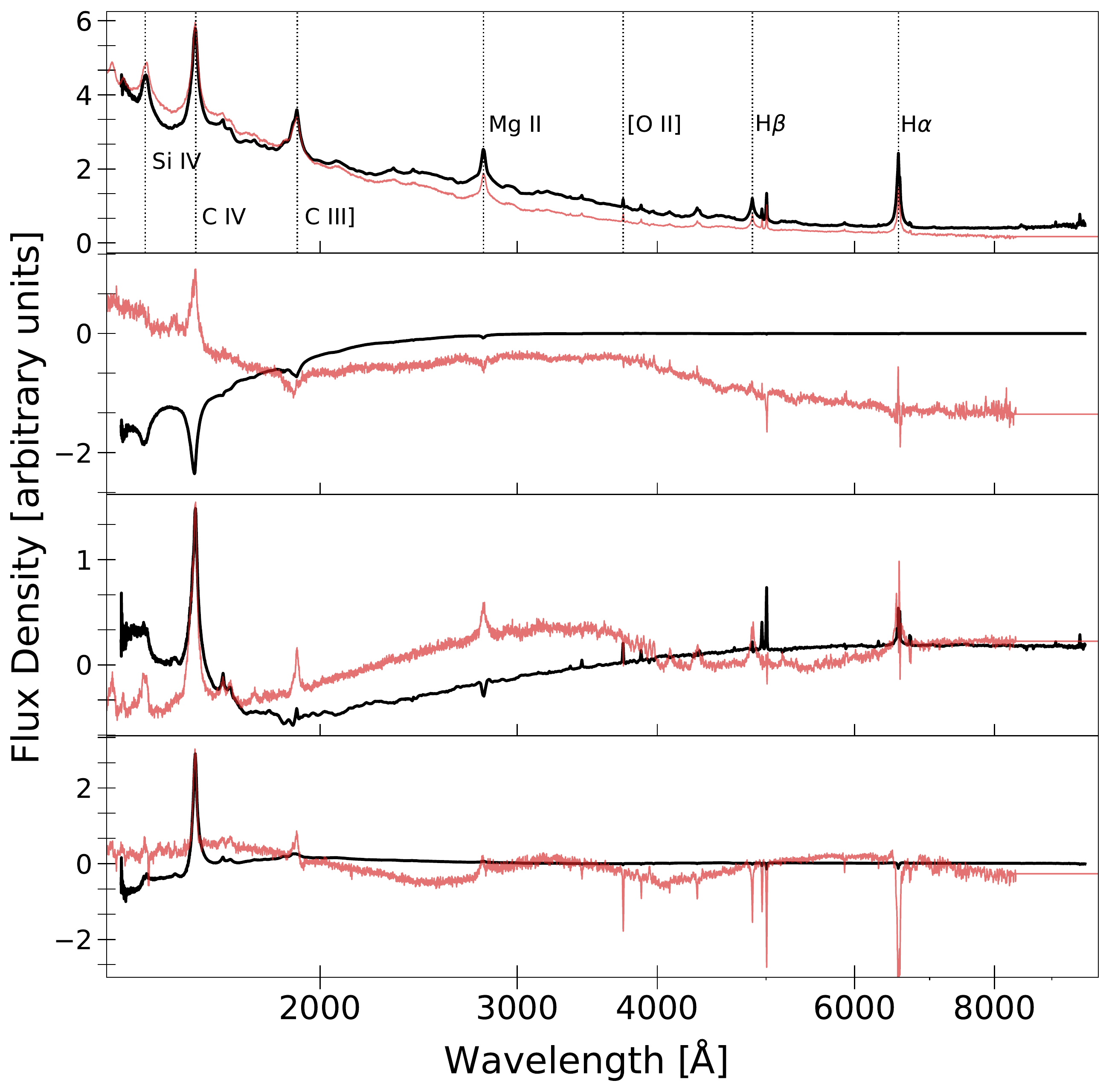}
	  \caption{The high redshift templates are shown on the left in black and cover $1.4<z<7.0$. The low 
	  redshift templates are shown on the right in black and cover $0.05<z<1.6$. The previous DESI QSO 
	  templates are overlaid in red in both figures. The previous templates are trimmed to the wavelength
	  coverage of the new templates and scaled for clarity in comparison. Prominent QSO 
	  emission lines are annotated on the mean spectra. The gray shading indicates the wavelength region 
	  that is poorly constrained in both eigenspectra sets owing to low QSO numbers and degraded 
	  spectrophotometric calibration at the shortest wavelengths of SDSS spectra.}
	  \label{fig:fig3}
\end{figure*}

We tested several redshift boundaries when defining the training subsamples for the high and low redshift QSO
eigenspectra. The maximum redshift for the low redshift training sample varied between $z=1.4$ to $z=1.9$ while 
the minimum redshift for the high redshift training sample varied between $z=1.0$ to $z=1.5$. The two training
samples always overlapped in redshift by $\Delta z = 0.4$ to allow a smooth transition from one set to the next
when classifying DESI spectra with Redrock. 

All high redshift templates are padded with a fast exponential decay beyond $\lambda \approx 600$~\AA\ to 
extend redshift coverage to $z=7.0$. Though QSO number density is low beyond $z \approx 4.5$ in DESI, the
padding ensures coverage for classifying the highest redshift observations. The choice of exponential decay is 
so that the flux at these wavelengths is close to zero, as expected beyond the Lyman break, while avoiding 
zeros and discontinuities. A single template set that was trained on all redshifts, analogous to the previous 
DESI QSO templates, was also created for comparison.

In all cases, the templates consisted of the mean spectrum of its training sample plus the first N 
eigenspectra from EMPCA on the residual spectra. N was varied for each template option to determine how many 
eigenspectra were necessary to achieve high-accuracy QSO classification on the DESI Truth Table without 
introducing problems in the galaxy sample from over-flexibility. We evaluated the accuracy, completeness, and 
catastrophic failure rates of classifications with Redrock on the DESI Truth Table (similar to 
Section~\ref{subsect:classification_accuracy}) as well as the quality of the fits to determine the optimal 
choice for classification in DESI. The full redshift range template set under-performed relative to all high and 
low redshift template combinations evaluated in this work for all N tested. 

We find that a mean plus three eigenspectra model in which the low redshift set, trained on $z < 1.7$ spectra,
classifies QSOs over $0.05<z<1.6$ and the high redshift set, trained on $z > 1.3$ spectra, classifies QSOs over 
$1.4<z<7.0$ performs the best. This choice of redshift boundaries outperformed other models by $\sim5-10$\% in 
terms of completeness. Including higher-order eigenspectra marginally improves completeness but at the expense 
of reduced galaxy sample completeness and quasar sample purity below $z=1.5$. The redshift range for 
classification differs from training due to the differences in wavelength coverage between SDSS and DESI and 
template binning in Redrock. We report only the results from the best-performing templates in 
Section~\ref{sect4:perftest}, omitting the performance of the alternative options for brevity. 

\begin{table*}[htb]
\centering
\caption{The performance metrics of Redrock alone and Redrock with the QSO afterburners on spectra from the
DESI Truth Table and QSO Main. The QSO Main is a subset of the DESI Truth Table mimicking main survey QSO
performance by limiting to QSO targets with $r<23$. The left section of the table lists the performance
metrics when using  the previous QSO templates and the right section is with the new QSO templates.}

\centerline{
\tablewidth{0pt}
\hskip-2.5cm\begin{tabular}{c | c c c c | c c c c} 

\hline
& \multicolumn{4}{c|}{\textbf{Previous QSO Templates}}  & \multicolumn{4}{c}{\textbf{New QSO Templates}}  \\
\hline
\multirow{2}{*}{Sample} & \multirow{2}{*}{Size} & \multirow{2}{*}{Contamination} & Catastrophic & \multirow{2}{*}{Incompleteness} & \multirow{2}{*}{Size} & \multirow{2}{*}{Contamination} & Catastrophic & \multirow{2}{*}{Incompleteness} \\
       &      &               & Failures     &                &      &               & Failures      &                \\
\hline
QSO      &  1,422   &  62 (4.4\%)   &  37 (2.6\%)   &  283 (17\%)   &  1,497   &  58 (3.9\%)   &  25 (1.7\%)   & 204 (12\%)   \\
QSO Main &  1,200   &  25 (2.1\%)   &  20 (1.7\%)   &  193 (14\%)   &  1,253   &  16 (1.3\%)   &  10 (0.8\%)   & 131 (9.6\%)  \\
Galaxy   &  15,310  &  491 (3.2\%)  &  268 (1.8\%)  &  172 (1.1\%)  &  15,237  &  431 (2.8\%)  &  258 (1.7\%)  & 185 (1.2\%)  \\

\hline
& \multicolumn{8}{c}{\textbf{with QSO afterburners}}  \\
\hline
QSO      &  1,781   &  216 (12\%)  &  77 (4.3\%)   &  78 (4.7\%)   &  1,779   &  202 (11\%)   &  58 (3.3\%)   & 66 (4.0\%)   \\
QSO Main &  1,387   &  64 (4.6\%)   &  31 (2.2\%)   &  45 (3.3\%)   &  1,384   &  48 (3.5\%)   &  16 (1.2\%)   & 32 (2.3\%)  \\
Galaxy   &  14,951  &  271 (1.8\%)  &  193 (1.3\%)  &  311 (2.1\%)  &  14,955  &  289 (1.9\%)  &  210 (1.4\%)  & 325 (2.2\%)  \\

\hline
\end{tabular}
}

\label{table_VImetrics}
\end{table*}

The new QSO template sets are shown in Figure~\ref{fig:fig3} along with previous DESI QSO templates for 
comparison. Spectrophotometric calibration is degraded at the shortest observer frame wavelengths in SDSS 
spectra \citep[see][]{margala16}. As such, both the previous and high redshift QSO templates suffer from 
reduced modeling power at the shortest wavelengths, namely beyond the Lyman limit, where SDSS QSO numbers
are sparse. On the red side of the Lyman limit, the most notable difference from the previous to the current 
templates is an overall reduction of random noise attributed to the increased training sample size.

Figure~\ref{fig:fig3} also highlights the improvement of using template sets specialized for specific redshift 
intervals. The high redshift set is overall similar to the previous templates, but there is a marked difference 
at low redshift. Particularly, the spectra lack the unphysical break in continuum present in the previous 
templates near 4000~\AA. This is likely a result of the greater flux density and high variance at shorter 
wavelengths dominating the previous templates and leading to unphysical continua at longer wavelengths to 
maintain orthogonality.

\section{Performance on DESI Spectra}
\label{sect4:perftest}

We test the performance of the new QSO templates in Redrock on the DESI spectral samples introduced in
Section~\ref{subsect:validationsample}. The results are presented alongside the performance of the previous
QSO templates to quantify improvements to QSO classification in DESI. DESI also uses the two QSO afterburner 
algorithms, QN and a broad Mg~\textsc{ii} line finder (described in Section~\ref{sect1:intro}), post-Redrock to 
improve QSO sample completeness. These algorithms are dependent on Redrock output, so we rerun the QSO 
afterburners on the spectral samples to assess how the new templates impact their performance. Additionally, we
evaluate the resulting galaxy samples, as a primary concern with improving the QSO templates is degrading the 
galaxy sample quality \citep[e.g.][]{guy23}.

\subsection{Classification Accuracy and Completeness}
\label{subsect:classification_accuracy}

\begin{figure*}
  \centering
	  \includegraphics[width=0.95\textwidth, angle=0]{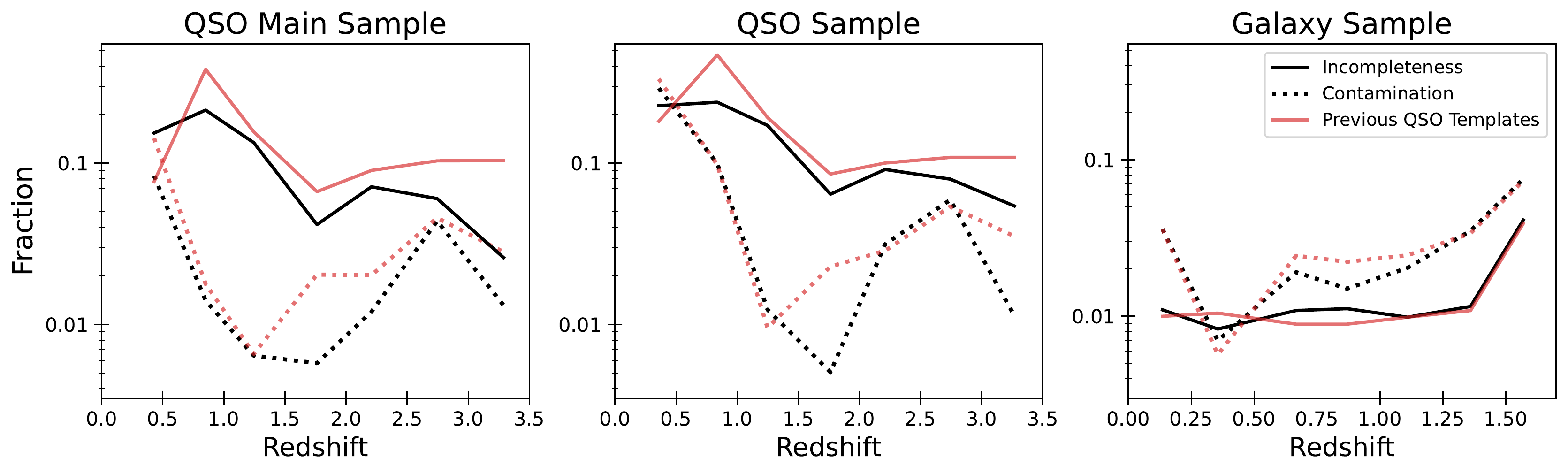}
	  
	  \caption{The contamination fraction (dashed) and incompleteness (solid) of the QSO Main sample, and of
	  the QSO and galaxy samples from the DESI Truth Table with redshift from Redrock with the new QSO 
	  templates. The same metrics for the previous QSO templates are shown in red. }
	  
	  \label{fig:fig4}
\end{figure*}

The VI labels in the DESI Truth Table and QSO Main samples include a spectral class and redshift, allowing us to 
estimate the contamination fraction, incompleteness, and catastrophic failure rates in the QSO samples from 
Redrock with the new QSO templates. We also review the Redrock galaxy sample from the DESI Truth Table for 
potential impacts from updating the QSO templates.

A contaminant in a sample is defined as a spectrum for which the VI and Redrock spectral classes disagree or the
Redrock redshift is bad. A good Redrock redshift for a QSO spectrum requires that 
\begin{equation}
    |\Delta v | = \frac{|z_{VI} - z_{Redrock}|}{1 + z_{VI}} \times c < 3000  \ensuremath{\, \mathrm{km~s}}^{-1} .
\end{equation}
We use a stricter definition of a good redshift for the galaxy sample, requiring  
$|\Delta v | < 1000$~km~s$^{-1}$. The catastrophic failure rate is a subset of the contamination fraction, 
isolating only bad Redrock redshifts regardless of spectral class correctness. The incompleteness of each sample
is the number of spectra with VI spectral class of QSO (galaxy) that is not included in the Redrock QSO (galaxy)
sample. 

The contamination fraction, catastrophic failure rate, and incompleteness for the Redrock QSO and galaxy 
samples from the DESI Truth Table and the Redrock QSO sample from QSO Main are reported in 
Table~\ref{table_VImetrics}. The same metrics for the previous QSO templates are also included in the table. 
Overall, both QSO samples show clear improvement with the change of templates while the galaxy sample exhibits 
minimal change. Assuming a binomial distribution, the new QSO templates lower the incompleteness of the QSO 
Main sample by $32\pm3$\% and reduce the catastrophic failure rate by $50\pm11$\%. A similar gain in 
completeness is achieved for the broader QSO sample as well, with a reduction in catastrophic failures of 
$32\pm8$\%.

Figure~\ref{fig:fig4} shows the contamination fraction and incompleteness with redshift for the three samples.
With the new templates, Redrock achieves a more complete QSO sample on the DESI Truth Table and QSO Main at all
redshifts $z>0.5$. This is reflected by the reduced contamination in the galaxy sample, indicating some QSOs
previously misclassified as galaxies are now being placed in the correct sample. At all redshifts, the new QSO
templates reduce the contamination in the QSO sample from QSO Main while achieving a lower or comparable
contamination fraction in the QSO sample from the DESI Truth Table.

\begin{figure*}
  \centering
	  \includegraphics[width=0.9\textwidth, angle=0]{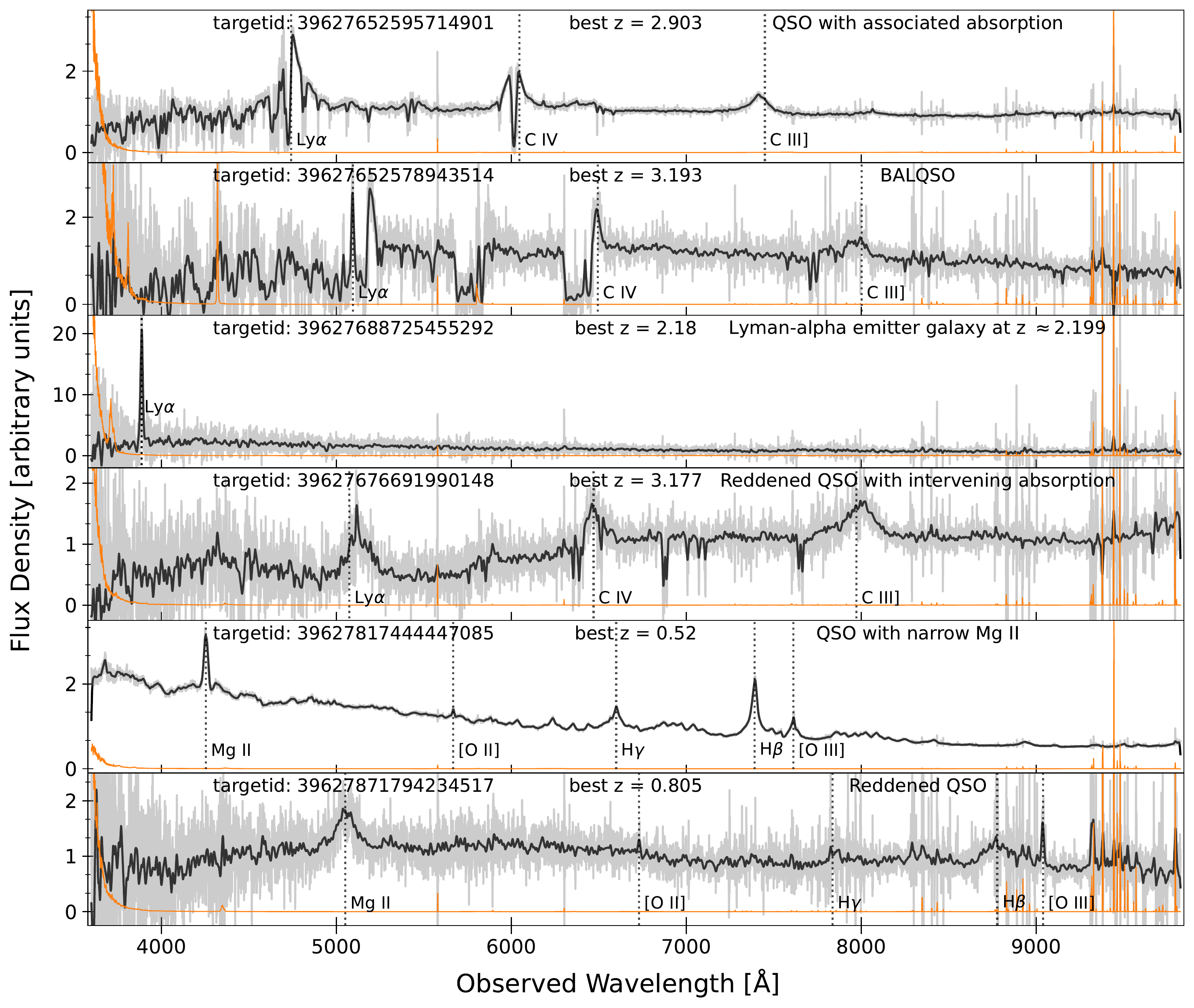}
	  
	  \caption{A subset of spectra from the DESI Truth Table that are classified as QSO by Redrock with the
	  new QSO templates, but not classified as QSO with the previous QSO templates (pre-QSO afterburners).
        The observed flux is shown in gray, the error on the flux is shown in orange, and the flux after 
        applying a Gaussian smoothing filter is shown in black. The DESI target ID, redshift estimate from the 
        new templates (best z), and basic information on the type of spectrum are annotated at the top of each
	  subplot. Prominent emission lines are indicated in the observer frame with vertical lines and labeled.}
	  
	  \label{fig:fig5}
\end{figure*}

There are 132 objects in the DESI Truth Table that are classified as QSO by Redrock with the new QSO templates
and were not classified as QSO with the previous templates. Figure~\ref{fig:fig5} shows a representative 
selection of these spectra, illustrating the types of QSO recovered. Many of the new Redrock QSOs display 
dust-reddened continua or strong absorption features that likely contributed to misclassification under the 
previous templates. A known issue with the previous templates is missing narrow or weak Mg~\textsc{ii} emission,
necessitating the Mg~\textsc{ii} QSO afterburner \citep{alexander23,chaussidon23}. The new templates recover
many of these as Redrock QSOs, though the afterburner is still necessary to reach the desired QSO completeness.

Figure~\ref{fig:fig5} also highlights a failure mode of Redrock generally. A small number of high redshift
Lyman-alpha emitter (LAE) galaxies are inadvertently targeted as QSO or galaxy. These systems are not 
classifiable as galaxy at the correct redshift with Redrock owing to the range of the galaxy templates 
($0<z<1.7$). As a result, many of these objects appear in the QSO sample, with the Ly$\alpha$ line flux fit by 
the QSO template Ly$\alpha$. The redshifts of true LAE classified as QSO are only approximate since Ly$\alpha$
emission is broad in a QSO spectrum. Future work in DESI may explore an intentional LAE target class with a 
devoted template set for classification and redshift fitting.

\vspace{1cm}

\subsubsection{Classification Accuracy and Completeness from Redrock with QSO Afterburners}
\label{subsubsect:classification_accuracy_QSOAFB}

The results reported so far are from Redrock alone. The QSO afterburners are necessary to achieve the required
classification performance for cosmological analyses. In this subsection, we evaluate the impact that the new
QSO templates have on classification for the full Redrock plus QSO afterburners pipeline. 

The contamination fraction, catastrophic failure rate, and incompleteness of the QSO and galaxy samples from 
Redrock with QSO afterburners on the DESI Truth Table are reported in the bottom half of 
Table~\ref{table_VImetrics}. As with the top half of the table, results are shown for both the previous and new
QSO templates. Also included in the table are the metrics for the QSO samples from QSO Main. 

The inclusion of the QSO afterburners results in an 8\% (12.3\%) gain in completeness for the QSO sample from 
the DESI Truth table and a 7.3\% (10.7\%) gain in completeness for QSO sample from QSO Main with the new 
(previous) templates. The contamination fraction of the QSO sample from the DESI Truth Table, however, increases
by about $7-8$\% with both template sets compared to Redrock alone. These QSO samples include all spectra
classified as QSO regardless of target type. The majority of the contamination ($\sim75$\% with previous and 
$\sim84$\% with new) is from spectra with a VI spectral class of `galaxy' for which there is strong agreement 
between the VI redshift and QSO template redshift. This type of contamination is inconsequential as the 
redshifting is successful. The remaining contamination results from incorrect redshift estimates compared to the 
VI redshift.

Considering only the Redrock plus QSO afterburners performance, the new QSO templates improve both QSO samples 
across all metrics while the galaxy sample is minimally affected. The QSO sample from QSO Main is the most 
reflective of main survey QSOs and therefore the Year 1 QSO sample owing to the target type and magnitude 
restriction. For this sample, the catastrophic failure rate is improved by $48\pm9$\%, the 
incompleteness by $29\pm7$\%, and the contamination by $25\pm5$\% using the new templates.

Overall, the new QSO templates produce QSO samples with higher completeness, less contamination, and a reduced
rate of catastrophic failures compared to the previous QSO templates. The improvement to the QSO classification 
performance is achieved with minimal impact on the galaxy sample.

\subsubsection{Average Spectral Properties of Redrock QSOs}

\begin{figure*}
  \centering
	  \includegraphics[width=0.85\textwidth, angle=0]{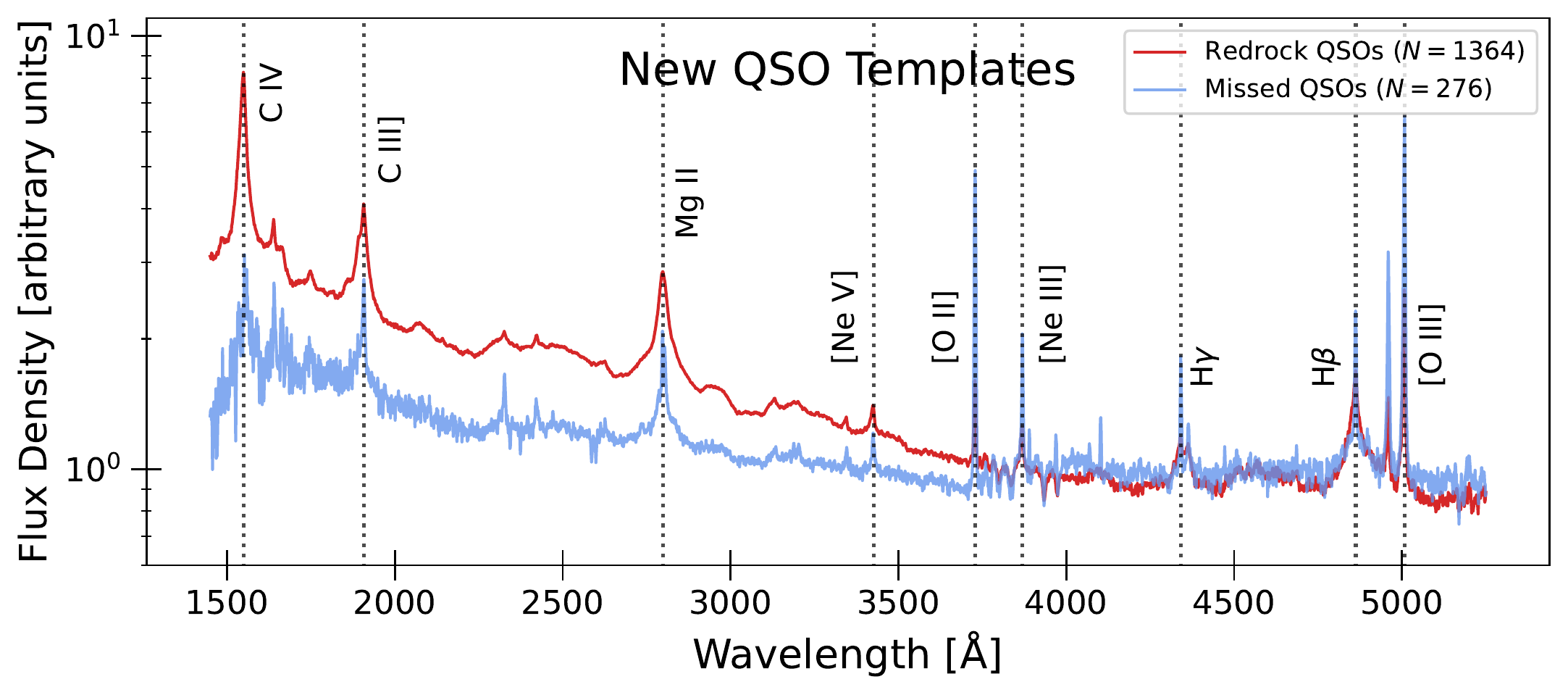}
	  \includegraphics[width=0.85\textwidth, angle=0]{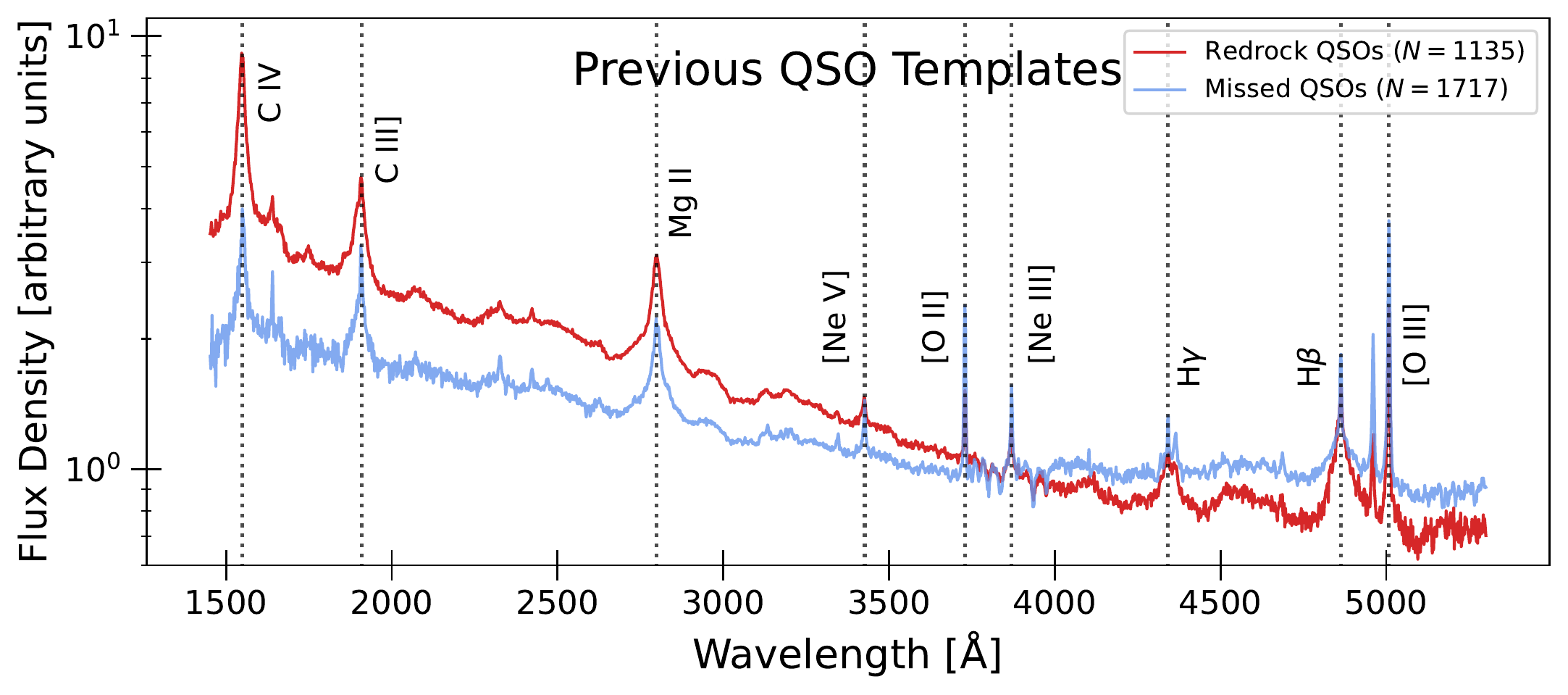}
	  
	  \caption{\textbf{Top:} Median composite of spectra in the DESI Truth Table classified as QSO by Redrock 
	  (red) and of spectra classified as QSO by the QSO afterburners but missed by Redrock (blue). The spectra
	  are normalized over $3800-3900$~\AA. Vertical lines indicate prominent QSO emission lines.
	  \textbf{Bottom:} Similar to above but with the previous QSO templates and an expanded sample of 
	  afterburner-identified QSOs (Figure~12 of \citet{alexander23}).}
	  
	  \label{fig:fig6}
\end{figure*}

The median composite of spectra in the DESI Truth Table classified as QSO by Redrock with the new QSO templates
is shown in the top panel of Figure~\ref{fig:fig6}. Also shown is the median composite of QSO spectra that were 
missed by Redrock but recovered by the QSO afterburners. This figure does not fully convey the diversity of 
Redrock and missed QSOs but illustrates the average spectral properties seen in a typical Redrock QSO spectrum 
while providing insight into the type of spectral features that the new QSO templates struggle to reconstruct. 

The bottom panel of Figure~\ref{fig:fig6} shows similarly constructed composite spectra for the previous QSO 
templates, originally presented by \citet{alexander23}. The missed QSO composite is constructed from a larger 
sample than with the new QSO templates, as their study focused on afterburner performance.

The typical spectrum classified as QSO by Redrock with the new QSO templates differs most from the typical
missed QSO spectrum at $\lambda_{RF} < 3700$~\AA\ where their underlying continua diverge. The QSOs missed by
the new QSO templates tend to display more reddened continua in this range, similar to those missed by the
previous QSO templates. For both template sets, a typical missed QSO may also display stronger host galaxy
features such as the Balmer series and Ca H+K stellar absorption lines observed over $3700-4000$~\AA\ as well
as increased [O~\textsc{ii}]and [O~\textsc{iii}] flux. While the afterburners recover many of the QSO spectra
that Redrock missed, a few percent of QSOs are still missed entirely (Table~\ref{table_vi_results}). Often,
these QSOs have low S/N, very red continua, or strong absorption features.

A notable difference in the missed QSO spectra from changing the QSO templates appears in the equivalent width
of broad lines such as C~\textsc{iv}, C~\textsc{iii}], and Mg~\textsc{ii}. The equivalent width of these lines
in the typical missed QSO is smaller with the new QSO templates than with the previous QSO templates. This
change indicates the new templates can describe more of the line profile parameter space, typically missing only
the most narrow of broad line systems.

\begin{table*}[htb]
\centering
\caption{Results from VI of spectra in Guadalupe that have discrepant Redrock-only classifications between the
previous and new QSO templates. We select two samples from each target type: redshift discrepant and spectral
class discrepant. Spectra in a given sample may exhibit both types of discrepancies however. Low Quality are
spectra that could not be confidently classified through VI.}

\centerline{
\tablewidth{0pt}
\hskip-2.5cm\begin{tabular}{c c c c c c c} 

\hline 
Target               & \multirow{2}{*}{Discrepancy} & Total     &  New Templates &  Previous Templates &  Neither  &  Low     \\
Type                 &                              & Inspected &  Correct       &  Correct            &  Correct  &  Quality \\

\hline
\multirow{2}{*}{QSO} & Redshift       &  100  & 70   &   16  &   5   &  9        \\
                     & Spectral Class &  100  & 70   &   8   &   12  &  10       \\
\hline
\multirow{2}{*}{ELG} & Redshift       &  100  &  48   &   9   &   20  &  23        \\
                     & Spectral Class &  100  &  61   &   9   &   5   &  25       \\
\hline
\multirow{2}{*}{LRG} & Redshift       &  90  &  28   &   9   &   28  &  25        \\
                     & Spectral Class &  100  &  59   &   0   &   33  &  8       \\
\hline
\multirow{2}{*}{BGS} & Redshift       &  48  &  13   &   4   &   14  &  17        \\
                     & Spectral Class &  100  &  72   &   1   &   24  &  3       \\

\hline
\end{tabular}
}

\label{table_vi_results}
\end{table*}

\subsubsection{Classification Discrepancies Between the Previous and New QSO Templates}
\label{subsubsect:classification_accuracy_MS}

As a last step in evaluating the catastrophic failures, completeness, and contamination, we visually inspect 
spectra in Guadalupe that have discrepant Redrock-only classifications between the previous and new QSO 
templates. Classification discrepancies occur when Redrock assigns a different spectral class (galaxy, QSO,
star) or redshift to the same spectrum depending on the QSO templates used. The threshold for a redshift to be
considered discrepant is defined as
\begin{equation}
    | \Delta v | = \frac{|z_{previous} - z_{new}|}{1 + z_{previous}} \times c > 3000  \ensuremath{\, \mathrm{km~s}}^{-1} .
\end{equation}
The breakdown of classification discrepancies by target type is shown in Table~\ref{table_class_change}.  

\begin{table}[htb]
\centering
\caption{Occurrences of classification discrepancies between the previous and new QSO templates by target type
in the Guadalupe sample}

\begin{tabular}{c | c c c c} 

\hline 
Target Type   &    QSO    &    ELG    &    LRG    &    BGS    \\
\hline
Total Targets         & 435,538 & 498,367 & 404,361 & 1,149,884 \\
Spectral Class        & 21,911  & 8,623   & 7,184   & 7,704     \\
Redshift              & 23,009  & 7,037   & 1,448   & 695       \\
Both                  & 13,326  & 4,394   & 1,254   & 497       \\

\hline
\end{tabular}
\label{table_class_change}
\end{table}

VI is a time-consuming task and inspecting all spectra in Guadalupe that change classification is infeasible. 
Instead, we VI only a subset of spectra from Table~\ref{table_class_change} to estimate the consequences of the
changed classifications. We choose 100 spectra that are discrepant in spectral class and 100 that have 
discrepant redshifts randomly from each target class over forty tiles. Exclusivity is not enforced between the
samples of discrepant spectra within a target type. Only 48 and 90 spectra have redshift disagreements over the
selected tiles for BGS and LRG targets, respectively. This leaves a final selection of 
738 spectra for VI.

Each spectrum in our VI sample was assigned to four inspectors who evaluated it using
Prospect\footnote{\href{https://github.com/desihub/prospect}{https://github.com/desihub/prospect}}. Prospect is
a VI tool developed for evaluating DESI's pipeline performance and target selection methods in the early survey. 
In short, Prospect displays an observed spectrum that can be interactively smoothed by the inspector. The top
nine best template fits from Redrock can be overlaid on the spectrum for reference. Properties associated with 
each fit are also available in the VI tool, including the $\chi^2$, redshift, and spectral class 
\citep[see][for detailed descriptions of Prospect]{lan23}. Each inspector indicates a spectral class and
redshift of the spectrum along with the confidence of their assessment on a scale of 0 to 4. A confidence of 0 
indicates no signal, 1 indicates there is one strong spectral feature present with a probable identification, 2
indicates more than one spectral feature has been identified, but features may be weak, and 4 indicates two or
more spectral features are securely identified. An average confidence of 2.5 is the minimum value to be
considered a confident classification. 

The VI results of the four inspectors inform a final classification ``truth'' table for the sample. Special 
attention is given to resolving the classification of spectra for which the inspectors disagree. We use only 
confident VI classifications in our analysis, reporting the remainder as ``Low Quality''.

The classification under a given QSO template set is considered correct if (1) the Redrock spectral class agrees
with the VI spectral class and (2) the Redrock redshift is less than $3000$~km~s$^{-1}$ from the VI redshift for
QSO targets or less than $1000$~km~s$^{-1}$ from the VI redshift for galaxy targets. A few instances exist where
both template sets satisfy these criteria on a given spectrum. For these cases, the template set that provides
the redshift estimate closest to the VI redshift is labeled as correct. Discrepant classifications for which
neither template set satisfy both criteria are reported under ``Neither Correct''. The results from VI of the 
discrepant spectra are summarized in Table~\ref{table_vi_results}, broken down by target type and type of
discrepancy. 

The confident VI classifications of both spectral class and redshift discrepant spectra indicate an improvement
with the new QSO templates for all target types. Redshift discrepant spectra with an LRG or BGS target type,
however, have a high occurrence rate of neither template set achieving the correct classification. In most 
cases, the failure stems from problematic reductions of the spectra, such as discontinuities in flux, that 
impact template fitting ability rather than a problem with QSO templates themselves. 

\subsection{Redshift Precision and Bias}
\label{subsect:redshift_precision}

This subsection is devoted to the redshift performance of final classifications (Redrock with QSO afterburners)
with the new QSO templates. We present estimates of redshift precision, the catastrophic failure rate, and
average redshift bias. We also briefly discuss redshift performance specific to BALQSOs.

\subsubsection{Redshift Precision Measured by Repeat Exposures}
\label{subsubsect:repeat_exposures}

The multiple spectra for QSO targets in the SV Repeat Exposures sample can be used to determine a lower bound
on the redshift precision and estimate the catastrophic failure rate at main survey depth. We measure the 
precision as the dispersion of the velocity offsets between pairs of repeat exposures. The dispersion is
calculated using the median absolute standard deviation (MAD) scaled by $1.4828 \times 2^{-\frac{1}{2}}$, 
following \citet{alexander23,lan23}. The catastrophic failure rate is defined as the percentage of these pairs
with an absolute velocity offset, scaled by $2^{-\frac{1}{2}}$ to account for the measurement error on each
redshift, over the typical failure threshold for QSOs of $3000$~km~s$^{-1}$ and a stricter threshold of 
$1000$~km~s$^{-1}$.

The measure of precision from repeat exposures provides only a lower-bound estimate on the true precision of QSO 
redshift estimates owing to the complex dynamics of QSO spectral features. Emission lines are often shifted non-
uniformly from the systemic redshift, impacting the ability of a template fitting software to extract the 
systemic redshift. Line dynamics that impact the redshift estimate obtained by the QSO templates will impact the 
redshift estimate in every exposure. The scaled MAD effectively estimates consistency or statistical precision.

\begin{figure}
  \centering
	  \includegraphics[width=0.45\textwidth, angle=0]{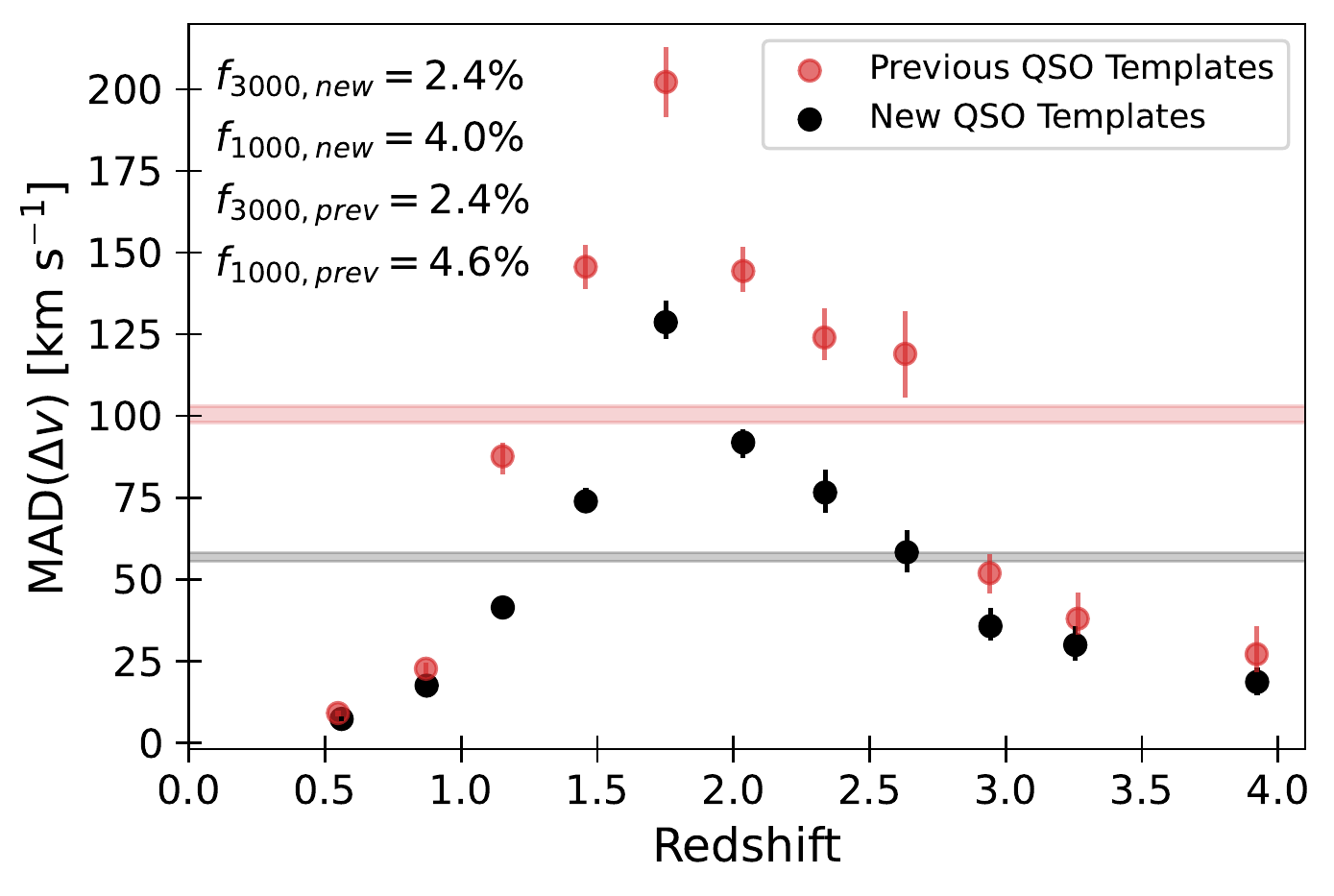}
	  
	  \caption{Redshift precision of QSO targets from the SV Repeat Exposures estimated by the scaled MAD 
        of the dispersion of velocity offsets between redshift pairs. The error bars indicate the 95\% confidence
	  interval, and the shaded regions show the redshift precision measured over all redshifts. $f_{1000}$ and
	  $f_{3000}$ are the catastrophic failure rate over the failure threshold of
	  $1000$~km~s$^{-1}$ and $3000$~km~s$^{-1}$, respectively.}
	  
	  \label{fig:fig7}
\end{figure}

The evolution of the scaled MAD with redshift is shown in Figure~\ref{fig:fig7} for both the new and previous
QSO templates along with their corresponding catastrophic failure rates. Both QSO template sets trend less 
precise as redshift increases until turning over at $z\approx1.75$, also reported by \citet{yu23}. The rise and 
fall of redshift precision between $1.35 \lesssim z \lesssim 2.5$ may be due to the presence of both the 
Mg~\textsc{ii} and C~\textsc{iv} emission lines in spectra. These lines have different expected velocity offsets 
from the systemic redshift which could introduce additional errors.

At main survey depth, the new templates clearly improve the precision at all redshifts. The redshift precision 
improves from $100^{+2.7}_{-2.4}$~km~s$^{-1}$ to $57^{+1.5}_{-1.3}$~km~s$^{-1}$ globally. The precision of 
both template sets is well below the DESI scientific and survey requirement for random error between individual
QSO redshift estimates of 750~km~s$^{-1}$ \citep{DESIcollaboration22}, even at the redshift where the 
uncertainty peaks. The new QSO templates do not appear to impact the catastrophic failure rate at the failure 
threshold of $3000$~km~s$^{-1}$. DESI scientific and survey requirements dictate that the rate of catastrophic 
failures exceeding $1000$~km~s$^{-1}$ shall be less than 5\% for tracer QSOs. Using this failure threshold on 
the SV Repeat Exposures sample yields a catastrophic failure rate of 4.0\% and 4.6\% for the new and previous 
QSO templates, respectively, meeting the requirements in both cases.

\subsubsection{Cross-correlation With Other Tracers}
\label{subsubsect:cross-correlation}
\begin{table*}
\centering
\caption{QSO redshift uncertainty measured from cross-correlations with LRGs, ELGs, and Ly$\alpha$ absorbers. 
$z_{\text{eff}}$ is the effective redshift of the correlation function, $\Delta r_{||}$ characterizes the 
average bias on QSO redshifts estimates, and $\sigma_{v}$ captures the effects of several redshift uncertainties 
including redshift error. A reduction of $\sigma_{v}$ demonstrates an improvement in the combination of 
statistical and systematic errors on the redshift estimates when using the new QSO templates.}

\centerline{
\tablewidth{0pt}
\hskip-2.5cm\begin{tabular}{c c | c c | c c | c}
\hline
&  & \multicolumn{2}{c|}{\textbf{Previous QSO Templates}}  & \multicolumn{2}{c|}{\textbf{New QSO Templates}}  \\
\hline
Redshift Range & Data Set &  $z_{\text{eff}}$ & $\Delta r_{||}$ [km~s$^{-1}$]  & $z_{\text{eff}}$ & $\Delta r_{||}$ [km~s$^{-1}$] & $\sigma_{v,new}$ / $\sigma_{v,prev}$  \\
 \hline
 
$0.8 < z < 1.1$ & LRG$\times$QSO       &  0.954  &  $-75 \pm 10$  &  0.954  &  $-3 \pm 10$   &  $0.990 \pm 0.038$ \\
$0.8 < z < 1.1$ & ELG$\times$QSO       &  0.961  &  $-40 \pm 11$  &  0.961  &  $31 \pm 8$    &  $0.986 \pm 0.046$ \\
$1.1 < z < 1.6$ & ELG$\times$QSO       &  1.353  &  $-17 \pm 8$   &  1.352  &  $59 \pm 8$    &  $0.921 \pm 0.035$ \\
$2.0 < z < 2.5$ & Ly$\alpha\times$QSO  &  2.171  &  $-132 \pm 31$ &  2.172  &  $9 \pm 29$   &  $0.948 \pm 0.307$ \\
$2.5 < z < 4.0$ & Ly$\alpha\times$QSO  &  2.753  &  $-236 \pm 46$ &  2.756  &  $-84 \pm 31$  &  $0.501 \pm 0.152$\\

\hline
\end{tabular}
}

\label{table_crosscorr}
\end{table*}
 
Bias and systematic uncertainty on QSO redshifts manifest in the characteristics of the small-scale peak of 
cross-correlation functions in the radial direction. The radial shift of the peak from zero, $\Delta r_{||}$,
quantifies the average bias on the QSO redshift estimates while the width of the peak, $\sigma_v$, is 
influenced by redshift precision \citep[][]{fontribera13,dMdB17,bault23}. An absolute value of QSO redshift 
precision cannot be determined from this measurement, however, as several cosmological parameters and 
non-linear effects contribute to $\sigma_{v}$ in addition to redshift error. We instead use the relative 
$\sigma_v$ between the correlation functions produced by the different QSO template sets with a given non-QSO
tracer to quantify the change in QSO redshift precision from changing templates. We assume these distortions of
the small-scale peak of the cross-correlation functions arise from the QSO redshift errors and not from the 
redshift errors on the other tracers. The physics of non-QSO tracer systems is better constrained, so we treat 
non-QSO redshifts as effectively error-free relative to the QSO redshifts.

We calculate the two-point cross-correlation function between QSOs and galaxies in the One-Percent Survey and
between QSOs and Ly$\alpha$ absorbers in Guadalupe using the 
pycorr\footnote{\href{https://github.com/cosmodesi/pycorr}{https://github.com/cosmodesi/pycorr}} software
package and the picca\footnote{\href{https://github.com/igmhub/picca}{https://github.com/igmhub/picca}} 
software package, respectively. All correlation functions are determined over radial separations of 
$-40 < r_{||} < 40 $ ~Mpc~h$^{-1}$ and integrated from 0 to 200~Mpc~h$^{-1}$ in the transverse direction.

\begin{figure}
  \centering
	  \includegraphics[width=0.45\textwidth, angle=0]{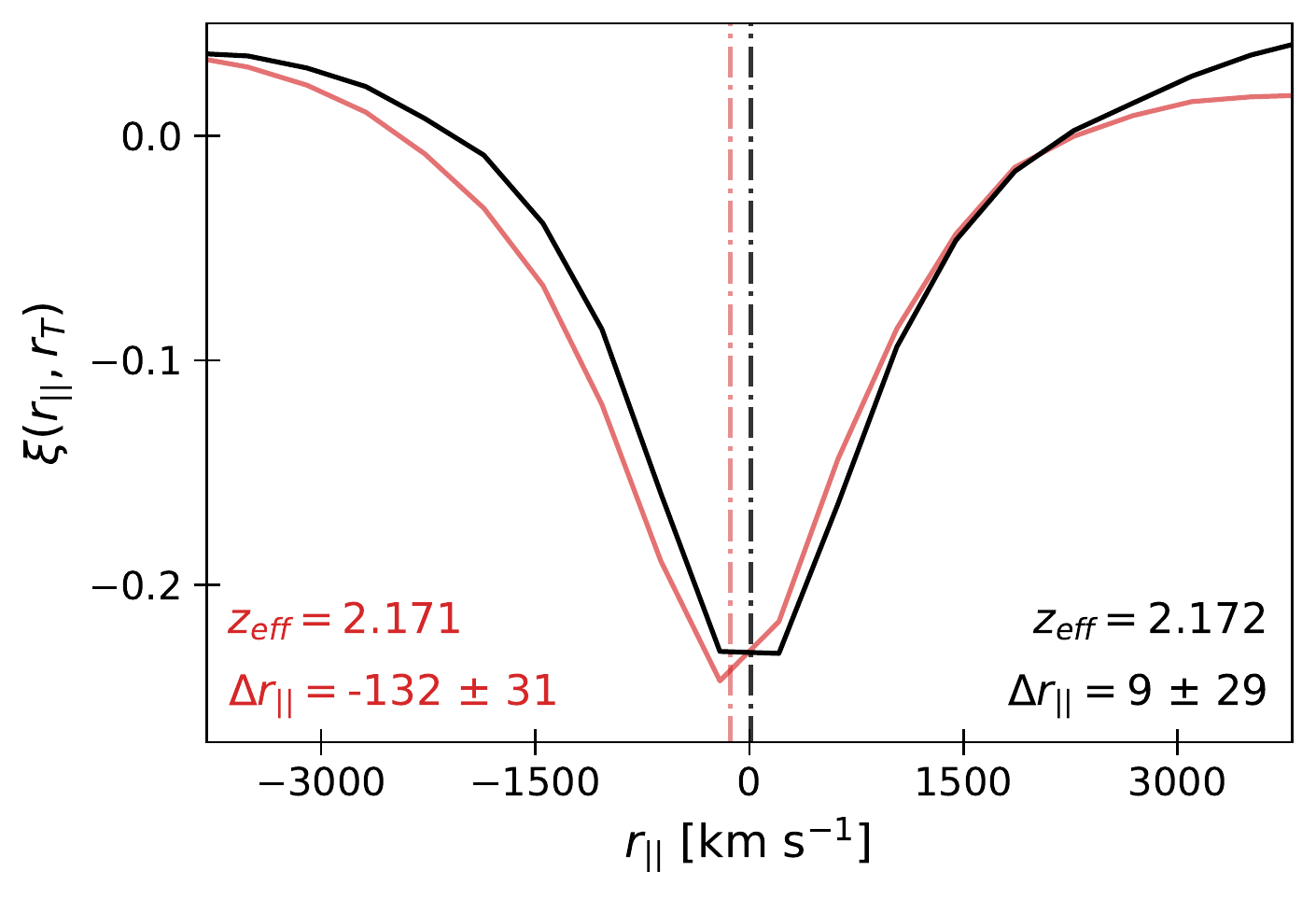}
	  
	  \caption{The Ly$\alpha$-QSO cross-correlation measured with QSOs at $2.0 < z < 2.5$ using the new QSO 
   templates (black) and the previous QSO templates (red). The best-fit $\Delta r_{||}$ for each measurement is 
   indicated with a vertical line. $z_{eff}$ is the effective redshift of the correlation function.}
	  
	  \label{fig:fig8}
\end{figure}

We determine the cross-correlation function between QSOs and LRGs over $0.8 < z < 1.1$, QSOs and ELGs over 
$0.8 < z < 1.6$, and QSOs with redshifts of $2.0 < z < 4.0$ and Ly$\alpha$ absorbers along their sight-lines. 
The Ly$\alpha$-QSO cross-correlation function is fit jointly with the Ly$\alpha$ auto-correlation to reduce 
parameter degeneracy. DESI ELG targets overlap with DESI QSO targets, so we require a `galaxy' classification
for ELGs in these measurements. The ELG and Ly$\alpha$ cross-correlation measurements are split on redshift to
check for any evolution in the bias and precision of the QSO redshifts.  

Figure~\ref{fig:fig8} shows an example of the small-scale signal we are measuring in the 
Ly$\alpha$-QSO cross-correlation. The x-axis corresponds to $|r_{||}| \lesssim 37$~Mpc~h$^{-1}$, converted to 
km~s$^{-1}$ for interpretation as redshift errors. The amplitude of the small-scale peak is negative because
Ly$\alpha$ is a negative bias tracer. The difference in peak width between the previous and new QSO templates 
seen in Figure~\ref{fig:fig8} corresponds to the difference in QSO redshift precision. This change in precision 
is captured by the ratio of  $\sigma_v$ in Table~\ref{table_crosscorr}.

We determine the best-fit value for the radial shift of the small-scale peak, $\Delta r_{||}$, and the 
corresponding peak width, $\sigma_v$, for each correlation function. The best fit $\Delta r_{||}$ 
values are given in Table~\ref{table_crosscorr} for each tracer population along with the ratio of $\sigma_v$ 
between correlation functions produced by the different QSO templates.

Redshifts from the new QSO templates are significantly less biased in most correlations.
$\Delta r_{||}$ is consistent with zero at 68\% confidence over two of the redshift ranges evaluated, a 
statement that is not true at any redshift with previous QSO templates. The exception to improvement in 
absolute bias comes from the ELG-QSO cross-correlation. The new QSO templates produce absolute bias consistent
with the previous QSO templates over $0.8<z<1.1$, but the bias over $1.1<z<1.6$ increases as a result of the
general 70~km~s$^{-1}$ shift seen in all galaxy-QSO cross-correlations. This shift can likely be attributed to 
the slight difference in rest frame between the previous and new template 
sets (Section~\ref{subsect:redshiftcorrection_comps}).

A large reduction in $\sigma_v$ is observed at the highest redshifts, indicating greatly improved redshift
precision. The value of $\sigma_v$ decreases with the new QSO templates across all tracers at the other
redshifts evaluated, indicating some improvement at $z<2.5$ though to a lesser extent. The observed 
improvement in precision is consistent with the findings of Section~\ref{subsubsect:repeat_exposures}.

\subsubsection{Redshift Success for BALQSOs}
\label{subsect:BAL_redshifts}

\begin{table*}
\centering
\caption{The average bias on and relative uncertainty between redshifts from the previous and new QSO 
templates for BALQSOs (AI~$>0$), as measured from cross-correlation with Ly$\alpha$ absorbers. We report 
$\Delta r_{||}$ for two correlation functions under each QSO template set: using the Redrock with QSO 
afterburner redshifts (no masking) and updated redshifts after masking BALs (masked). Additionally, relative 
uncertainty (relative $\sigma_v$)  between the two correlation functions under a given template set is also 
reported, measuring the impact of masking BALs on redshift uncertainty.}

\centerline{
\tablewidth{0pt}
\hskip-2.5cm\begin{tabular}{c | c c c | c c c | c }
\hline
& \multicolumn{3}{c|}{\textbf{Previous QSO Templates}}  & \multicolumn{3}{c|}{\textbf{New QSO Templates}}  & \\
\hline
Data Set &  $z_{\text{eff}}$ & $\Delta r_{||}$ [km~s$^{-1}$]  & $\sigma_{v,mask}$ / $\sigma_{v,no~mask}$ & $z_{\text{eff}}$ & $\Delta r_{||}$ [km~s$^{-1}$] & $\sigma_{v,mask}$ / $\sigma_{v,no~mask}$  & $\sigma_{v,new}$ / $\sigma_{v,prev}$  \\
 \hline

Ly$\alpha\times$BALQSO & \multirow{2}{*}{2.402}  &  \multirow{2}{*}{$-254 \pm 60$} & \multirow{4}{*}{$0.527 \pm 0.213$} & \multirow{2}{*}{2.413} &  \multirow{2}{*}{$-64 \pm 43 $}  &  \multirow{4}{*}{$0.643 \pm 0.288$} & \multirow{2}{*}{$0.584 \pm 0.261$} \\ 
(masking) & & & & & & & \\
Ly$\alpha\times$BALQSO & \multirow{2}{*}{2.403}  &  \multirow{2}{*}{$-177 \pm 63$} &                    & \multirow{2}{*}{2.414}  & \multirow{2}{*}{$56 \pm 47$}  &  & \multirow{2}{*}{$ 0.478 \pm 0.193$} \\ 
(no masking) & & & & & & & \\



\hline

\end{tabular}
}

\label{table_BALcrosscorr}
\end{table*}

BALQSOs are notorious for degraded redshift success relative to non-BALQSOs. The presence of BALs increases the
catastrophic failure rate as well as the random and systematic redshift errors of the total QSO sample. A PCA
template fitting algorithm that uses the eigenspectra of \citet{guo19} was developed to classify high 
ionization BALQSOs in the DESI QSO sample over $1.57<z<5.0$. The classifier records the location and width of
BALs associated with the C~\textsc{iv} and Si~\textsc{iv} emission lines. In this work, we define a BALQSO as
having an intrinsic absorption index \citep[AI;][]{hall02} relative to C\textsc{iv} that is greater than zero. 

Refined redshifts for BALQSOs in Guadalupe are provided in the DESI EDR BAL catalog presented by 
\citep{filbert23}. The authors refit affected spectra with the Redrock QSO templates after masking BAL 
features identified by the BAL classifier. The updated fit can vary up to $| \Delta z | = 0.1$ from
the initial redshift estimate given by Redrock with the QSO afterburners. We repeat this re-redshifting 
procedure for BALQSOs in Guadalupe when the new QSO templates are used in Redrock.

We compute the cross-correlation between BALQSOs (AI $> 0$) in Guadalupe and the Ly$\alpha$ absorber sample
from the previous subsection. As in Section~\ref{subsubsect:cross-correlation}, we determine the average bias
on BALQSO redshifts and the relative redshift uncertainty between the previous and new QSO template 
sets. We compute the cross-correlations using both the initial redshift estimates from Redrock with QSO 
afterburners and the updated redshifts after masking BALs, leading to a total of four cross-correlation 
measurements.

The best fit $\Delta r_{||}$ value for each correlation function is given in Table~\ref{table_BALcrosscorr}  
along with relative $\sigma_{v}$ between correlation functions produced by the different template sets. The 
new QSO templates significantly improve BALQSO redshift performance. The average absolute bias on both the
initial and masked redshifts is reduced by over $100$~km~s$^{-1}$. Additionally, the redshift uncertainty
improves by an estimated $40-50\%$. 

The relative $\sigma_{v}$ between the correlation functions produced by the masked and initial redshifts under
each template set are also reported in Table~\ref{table_BALcrosscorr}. For both template sets, the masking 
procedure greatly improves the redshift uncertainty for BALQSOs while shifting the average bias toward more 
negative velocity offsets. These findings agrees with \citet{garcia23} in that masking BAL features improves
redshift errors and with \citet{filbert23} that it results in lower redshifts for BALQSOs on average.

\section{Summary}
\label{sect5:summary}

The new DESI QSO templates are a significant update to the automatic spectral classification procedure in
DESI. The previous DESI QSO templates, inherited from (e)BOSS, were designed to classify QSO spectra across
the full range of $0.05<z<7.0$. This broad redshift coverage limits the ability to adequately reconstruct the
full range of spectral diversity, contributing to poor completeness when used as the sole method for 
identifying QSO spectra in DESI. We have developed new QSO templates for DESI consisting of a low and high 
redshift set trained to classify QSOs over $0.05<z<1.6$ and $1.4<z<7.0$, respectively. We have shown that 
these QSO templates produce more complete QSO samples while providing higher accuracy redshift estimates.

The new QSO templates were trained on approximately 200,000 (e)BOSS QSO spectra over $0.0 < z < 5.0$. Clusters 
of similar spectra were identified using the clustering technique of B22 and combined into high S/N composite 
spectra. The enhanced S/N enabled direct fitting of the Mg~\textsc{ii} emission line, a relatively stable 
emission feature with low bias relative to the systemic redshift. The wavelength solution of all composite 
spectra were corrected so that the Mg~\textsc{ii} line was at its proper rest frame. Eigenspectra of the
redshift-corrected composite spectra were used to rederive redshifts for the individual spectra of the 
training sample, placing all spectra in a consistent rest frame. Poor fits from the eigenspectra indicated
problematic and misclassified spectra in the training sample, facilitating their removal. The high and low
redshift QSO template sets that comprise the final QSO templates for DESI were derived from PCA on subsets of
the updated training sample in redshift space. The redshift boundaries of the high and low redshift template 
sets were determined through maximizing the completeness and redshift success of Redrock on the DESI Truth 
Table. 

We compared the classifications from Redrock with the new QSO templates and with the previous QSO templates on 
the DESI Truth Table and the QSO Main sample. Both samples include spectra that are exposed beyond main survey 
depth, facilitating classification via visual inspection. The QSO Main sample is a subset of the DESI Truth 
Table limited to QSO targets that are representative of DESI main survey QSO targets ($r<23$). For this 
sample, the catastrophic failure rate is reduced by approximately half with the new templates while the QSO 
incompleteness problem with Redrock is mitigated.

DESI's two QSO afterburner algorithms work in conjunction with Redrock to achieve the required QSO
completeness for scientific analyses. The new QSO templates provide less contaminated and more complete QSO
samples from the Redrock plus QSO afterburners classification scheme than the previous QSO templates. The 
improvements to QSO classification are achieved with minimal consequences observed in the galaxy sample.

The most significant impact of the new QSO templates is observed in the redshift precision and accuracy. The
velocity dispersion of repeat exposure redshifts in the SV Repeat Exposures sample indicates that the 
statistical redshift precision of QSOs at main survey depth is improved by $40-50$\% globally with no apparent
change in the catastrophic failure rate at the $3000$~km~s$^{-1}$ threshold. The improvement in precision is
largest over $1.0<z<2.75$. The small-scale cross-correlation of QSOs with other DESI tracers reveals a
reduction in absolute bias on QSO redshifts in most measurements, though a redshift dependency in
the bias persists (Table~\ref{table_crosscorr}). We also measure a significant reduction in the
absolute bias on the redshift estimates for BALQSOs while improving their redshift uncertainty by an estimated
$40-50$\% (Table~\ref{table_BALcrosscorr}). In future template versions, further improvement in 
redshift performance may be achieved with more complete modeling of the Mg~\textsc{ii} emission line in the 
training sample spectra.

Future work will explore potential solutions to reduce or remove the redshift-dependent bias on QSO redshifts
at the highest redshifts. One possible method to reduce the bias and mitigate its redshift dependency is 
through correcting for the optical depth of the Ly$\alpha$ forest while determining the redshift. This would 
require a new set of high redshift templates trained with a model of the Ly$\alpha$ forest optical 
depth evolution folded in. These upgraded high redshift templates could be constructed with DESI QSOs 
to better capture spectral diversity specific to DESI QSO targets. Incorporating the Ly$\alpha$ forest optical 
depth evolution into a template set would also allow predictions of the average continuum for unbiased 
estimates of the Ly$\alpha$ absorption field. Such continuum predictions would allow modeling of the density 
field on all scales for cosmological studies, including determining clustering amplitude through cross-
correlations with the cosmic microwave background. A more aggressive approach to reducing bias is masking 
pixels in the Ly$\alpha$ forest during redshift determination. This option is at the sacrifice of information 
in the forest, though the typical S/N of forest pixels is much less than that in the rest of the spectrum.

\subsection{DESI Science Requirements}

The scientific and survey requirements for DESI as a Stage~IV dark energy survey are detailed in Table~3 of
\citet{DESIcollaboration22}. Below, we map the performance of the new QSO templates to the relevant scientific
and survey requirements for redshift success of tracer ($z<2.1$) and Ly$\alpha$ ($z>2.1$) QSOs.

\vspace{0.25cm}

\noindent \textbf{L2.4.2, L2.5.2:} \emph{The random redshift error shall be less than 
$\sigma_{v} = 0.0025(1 + z)$, or equivalently $750$~km~s$^{-1}$ rms.} 

\noindent The median absolute standard deviation (MAD) of velocity offsets between redshift pairs in the SV
Repeat Exposures sample measures the statistical precision, or the random redshift error, of
redshift estimates for QSO targets in DESI. As shown in Figure~\ref{fig:fig7}, the statistical redshift
precision of the new QSO templates is less than $150$~km~s$^{-1}$ over $0.0<z<4.0$ and drops below
$50$~km~s$^{-1}$ for $z < 1.25$ and $z > 2.75$.

\vspace{0.25cm}

\noindent \textbf{L2.4.3:} \emph{Systematic inaccuracy in the mean redshift of tracer QSOs shall be less than 
$\Delta z = 0.0004(1 + z)$ or equivalently, $120$~km~s$^{-1}$.} 

\noindent The average inaccuracy, or bias, on QSO redshifts is measured as $\Delta r_{||}$ in the fit to
cross-correlation functions with other DESI tracers (Section~\ref{subsubsect:cross-correlation}). We only
measure the bias on tracer QSO redshifts over $0.8<z<1.6$, defined by the overlap in redshift distributions
with the other non-QSO tracers targeted by DESI. The maximum average inaccuracy in this range is 
$59 \pm 10 $~km~s$^{-1}$. Additionally, the average inaccuracy of Ly$\alpha$ QSO and BALQSO redshifts over
$2.0 < z < 4.0$ are less than $120$~km~s$^{-1}$ with the new templates, though no minimum redshift 
accuracy is specified for high redshifts.

\vspace{0.25cm}

\noindent \textbf{L2.4.4:} \emph{Catastrophic failures exceeding $1000$~km~s$^{-1}$ shall be less than 5\% for
tracer QSOs.}

\noindent The typical tracer QSO in DESI will receive an exposure equivalent to main survey depth. Therefore, the 
projected catastrophic failure rate is adequately estimated by limiting the SV Repeat Exposures sample to 
$z<2.1$. This yields a catastrophic failure rate of 3.9\% for tracer QSOs at the $1000$~km~s$^{-1}$ threshold.

\vspace{0.25cm}

\noindent \textbf{L2.5.3:} \emph{Catastrophic failures shall be less than 2\% for Ly$\alpha$ QSOs.}

\noindent Similar to above, a catastrophic failure rate for Ly$\alpha$ QSOs is estimated by limiting the SV
Repeat Exposures sample to $z>2.1$. The rate of catastrophic failures exceeding $1000$~km~s$^{-1}$ is 4.1\%
with the new QSO templates, compared to 5.6\% with the previous QSO templates. The new QSO templates achieve a
catastrophic failure rate of 1.7\% if the failure threshold is relaxed to $3000$~km~s$^{-1}$, compared to
1.9\% with the previous templates. Ly$\alpha$ QSOs, unlike tracer QSOs, will be observed at least four times
to enhance the S/N of their spectra. As such, the catastrophic failure rates determined here may be an
overestimate.

\section*{ACKNOWLEDGEMENTS}
{The work of Allyson Brodzeller and Kyle Dawson was supported in part by U.S. Department
of Energy, Office of Science, Office of High Energy Physics, under Award No. DESC0009959.

This research is supported by the Director, Office of Science, Office of High Energy
Physics of the U.S. Department of Energy under Contract No. DE–AC02–05CH11231, and by the
National Energy Research Scientific Computing Center, a DOE Office of Science User
Facility under the same contract; additional support for DESI is provided by the U.S. 
National Science Foundation, Division of Astronomical Sciences under Contract No. 
AST-0950945to the NSF’s National Optical-Infrared Astronomy Research Laboratory; the 
Science and Technologies Facilities Council of the United Kingdom; the Gordon and Betty
Moore Foundation; the Heising-Simons Foundation; the French Alternative Energies and Atomic 
Energy Commission (CEA); the National Council of Science and Technology of Mexico (CONACYT);
the Ministry of Science and Innovation of Spain (MICINN), and by the DESI Member Institutions:
\url{https://www.desi.lbl.gov/collaborating-institutions}.

The authors are honored to be permitted to conduct scientific research on Iolkam Du’ag (Kitt Peak),
a mountain with particular significance to the Tohono O’odham Nation.

Funding for the Sloan Digital Sky Survey IV has been provided by the Alfred P. Sloan
Foundation, the U.S. Department of Energy Office of Science, and the Participating 
Institutions. 

SDSS-IV acknowledges support and resources from the Center for High Performance 
Computing  at the University of Utah. The SDSS website is www.sdss.org.

SDSS-IV is managed by the Astrophysical Research Consortium for the Participating
Institutions of the SDSS Collaboration including the Brazilian Participation Group, 
the Carnegie Institution for Science, Carnegie Mellon University, Center for 
Astrophysics | Harvard \& Smithsonian, the Chilean Participation Group, the French
Participation Group, Instituto de Astrof\'isica de Canarias, The Johns Hopkins 
University, Kavli Institute for the Physics and Mathematics of the Universe 
(IPMU) / University of Tokyo, the Korean Participation Group, Lawrence Berkeley 
National Laboratory, Leibniz Institut f\"ur Astrophysik Potsdam (AIP),  
Max-Planck-Institut f\"ur Astronomie (MPIA Heidelberg), Max-Planck-Institut f\"ur 
Astrophysik (MPA Garching), Max-Planck-Institut f\"ur Extraterrestrische Physik 
(MPE), National Astronomical Observatories of China, New Mexico State University, 
New York University, University of Notre Dame, Observat\'ario Nacional / MCTI, 
The Ohio State University, Pennsylvania State University, Shanghai Astronomical 
Observatory, United Kingdom Participation Group, Universidad Nacional Aut\'onoma 
de M\'exico, University of Arizona, University of Colorado Boulder, University of 
Oxford, University of Portsmouth, University of Utah, University of Virginia, 
University of Washington, University of Wisconsin, Vanderbilt University, 
and Yale University.
}

\section*{DATA AVAILABILITY}
All data shown in figures are available at \href{https://doi.org/10.5281/zenodo.7872747}{https://doi.org/10.5281/zenodo.7872747}

\bibliographystyle{aasjournal}
\bibliography{references}

\end{document}